\documentstyle[12pt,psfig,aaspp4]{article}
\font\cap=cmcsc10

\def\ni{\noindent}        

\def\hi{\noindent \hangindent=2.5em}

\def\cm{{\rm\,cm}}
\def\kpc{{\rm\,kpc}}

\def\kmsec{{\rm\,km/s}}
\def\kms{\kmsec}
\def\hnot{{\rm\,km/s/Mpc}}
\def\msun{{\rm\,M_\odot}}
\def\lsun{{\rm\,L_\odot}}

\def\surfb{{\rm\,mag/arcsec^2}}

\def\araa{{\rm Ann.\ Rev.\ Astr.\ Ap.}, }
\def\aj{{\rm A.~J.}, }  
\def\apj{{\rm Ap.~J.}, }  
\def\apjs{{\rm Ap.~J.~Suppl.}, }  
\def\apjl{{\rm Ap.~J.~(Letters)}, } 
\def\fcp{{\rm Fund.~Cos.~Phys.}, } 
\def\mn{{\rm M.N.R.A.S.}, }      
\def\nat{{\rm Nature}, }      
\def\aa{{\rm Astr.~Ap.}, }     
\def\aasup{{\rm Astr.~Ap.~Suppl.}, }     

\def\spose#1{\hbox to 0pt{#1\hss}}
\def\lta{\mathrel{\spose{\lower 3pt\hbox{$\mathchar"218$}}
     \raise 2.0pt\hbox{$\mathchar"13C$}}}
\def\gta{\mathrel{\spose{\lower 3pt\hbox{$\mathchar"218$}}
     \raise 2.0pt\hbox{$\mathchar"13E$}}}

\def\clock{\count0=\time \divide\count0 by 60
     \count1=\count0 \multiply\count1 by -60 \advance\count1 by \time
     \number\count0:\ifnum\count1<10{0\number\count1}\else\number\count1\fi}

\begin{document}

\title{The Formation of Disk Galaxies}

\author{Julianne J. Dalcanton\altaffilmark{1,2}}
\affil{Observatories of the Carnegie Institution
	of Washington, 813 Santa Barbara Street, Pasadena CA, 91101 \\ \& \\
	Princeton University Observatory,
       Princeton, NJ 08544 }

\author{David N.\ Spergel, \& F J Summers\altaffilmark{3}}
\affil{Princeton University Observatory,
       Princeton, NJ 08544}

\altaffiltext{1}{e-mail address: jd@ociw.edu}
\altaffiltext{2}{Hubble Fellow}
\altaffiltext{3}{Current Address: Columbia Astrophysics Lab, Mail Code 5247,
550 West 120th Street, New York, NY 10027}
  
\begin{abstract}

We present a scenario for the formation of disks which explains not
only the properties of normal galaxies, but the properties of the
population of low surface brightness galaxies (LSBs) as well.  We use
a gravitationally self-consistent model for disk collapse to calculate
the observable properties of disk galaxies as a function of mass and
angular momentum of the initial protogalaxy.  The model naturally
produces both smooth, asymptotically flat rotation curves and
exponential surface brightness profiles over many disk scale lengths.

In this scenario, low mass and/or high angular momentum halos naturally
form low baryonic surface density disks, which will tend to be low
surface brightness.  Theoretical and numerical calculations suggest
galaxy halos should form with a wide range of mass and angular
momenta, and thus, the disks which form within these halos should have
a wide range of surface brightnesses and scale lengths.  We use
theoretical predictions for the distribution of halo masses and
angular momenta to explicitly calculate the expected number density of
disk galaxies as a function of central surface brightness and disk
scale length.  The resulting distribution is compared to the observed
properties of galactic disks, and is shown to explain the range of
observed disk properties, including the cutoff in the maximum disk
scale length as a function of surface brightness.  We also show that
disk instabilities explain the observed lack of high surface density
disks.  The calculated distribution of disk properties also suggests
that there are large numbers of galaxies which remain undetected due
to biases against galaxies with either low surface brightness or small
scale length.  We quantify this by calculating the difference between
the intrinsic luminosity function and the luminosity function which
would be measured in a galaxy survey with a given limiting surface
brightness.  We show that current measurements of the galaxy
luminosity function may be missing more than half of all $L_*$
galaxies, and an even larger fraction of faint galaxies, given the
correlation between mass and surface brightness.  The likely
underestimate of the luminosity density is also expected to be large.
We discuss how this affects observations of the ``faint blue galaxy''
population.

We also investigate the dynamics of galaxies as a function of surface
brightness.  We show that, in the absence of any systematic change in
the ratio of disk mass to disk luminosity, galaxies of all surface
brightnesses should lie on the same Tully-Fisher (1977) relation.  Our
models also show systematic changes in the shape of the rotation curve
as a function of angular momentum, which leads to low surface
brightness galaxies having slowly rising rotation curves.
Furthermore, because high angular momentum LSB disks have their
baryonic mass spread over a larger area than normal galaxies of
comparable mass, LSB disks contribute very little to the observed
dynamics of the galaxy.  Thus, LSBs provide a very effective tracer of
the shape and mass profile of the dark matter halo, out to
proportionally larger radii than is possible to observe with normal
galaxy rotation curves.

\end{abstract}

\singlespace

\section{Introduction}

Disk galaxies, including low surface brightness galaxies (LSBs), show
remarkable regularities: most disk galaxies have asymptotically flat
rotation curves (see Bahcall and Casertano 1985, Burstein \& Rubin 1985,
Persic et al. 1995, de Blok et al.\ 1996, and the  references
therein); the Tully-Fisher relation between circular velocity and
luminosity appears to hold over a wide range of both mass and surface
brightness (Strauss and Willick 1995, Zwaan et
al. 1995, Sprayberry 1995); and the light profile of galactic disks
are well fit by an exponential over several disk scale lengths (de
Jong 1995) for high surface brightness spirals, as well as for
non-dwarf LSBs (Davies, Phillips, \& Disney 1990, McGaugh \& Bothun
1994, James 1994, McGaugh et al 1995, Dalcanton et al.\ 1997, DeJong 1995).
The consistency of these properties suggests that the basic
structure of all disk galaxies is a general property of galaxy
formation, either tied to regulated feedback during formation, or, as
we shall consider in this paper, to primordial properties of
protogalaxies, namely their mass and angular momentum.  
The fact that galaxies' rotation curves are smooth in both the
disk-dominated inner regions and the halo-dominated outer regions
strongly suggests that the structure of the inner halo is coupled to
the formation of the disk.

Past disk formation models have been able to reproduce many of the
universal properties of disk galaxies (Fall \& Efstathiou 1980, Gunn
1982, van der Kruit 1987).  The models were motivated by Mestel's
(1963) observation that the angular momentum distribution of the
galactic disk is very similar to that of a sphere in solid body
rotation.  Thus, the models assumed that the collapse of a uniformly
rotating gaseous protogalaxy was a good model for the formation of the
disk.  In general, the models produced realistic disk surface density
profiles, provided that the models included the further assumptions
that 1) there was very little angular momentum transport while the gas
is collapsing, 2) the angular momentum was responsible for eventually
halting the collapse, and 3) the density profile of the dark matter
halo was such that the resulting galaxy will have a flat rotation
curve.  However, because of the third assumption, the resulting mass
distribution of the disk+halo system was not necessarily self
consistent; because the halo did not respond to the
mass of the collapsed disk, the final disk was not necessarily in
gravitational equilibrium.

One limitation of past models of disk formation is that they
have only attempted to
reproduce the properties of ``normal'', high surface brightness disks
such as our own.  However, a growing body of literature on low surface
brightness galaxies (LSBs) has revealed that there is a substantial
population of disk galaxies whose photometric properties fall well
outside what is usually classified as ``normal''.  Any viable
formation scenario must explain the tremendous range of disk galaxy
surface brightnesses and scale lengths, while simultaneously
explaining the tremendous similarities among disk galaxies throughout
this range.

In this paper we develop a gravitationally self-consistent scenario
for the formation of disks [\S\ref{formation}].  It is remarkably
effective at explaining both the structural and dynamical regularities
among disks, as well as for motivating the enormous range of galaxy
properties which are observed.  We use the model to calculate directly
observable properties of individual galaxies (light profiles
[\S\ref{surfdensect}], rotation curves [\S\ref{rotcurvesect}], etc.).
We also use it to compute the expected distribution of galaxies as a
function of surface brightness and disk scale length; this allows us
to easily calculate the degree to which selection effects shape the
observed properties of galaxies [\S\ref{surfbriden}].


\section{The Formation of Disks}		\label{formation}

The outline of the scenario which we will use for disk formation is as follows:

(1) In the early universe, tidal torquing spins up both the dark
matter and the baryonic matter in some region of space, which we will
approximate as an isolated sphere of radius $a$ in solid body rotation
(i.e. a rotating, spherical top-hat overdensity).  While the region
has enough of a mass overdensity that it will eventually collapse into
a galaxy, the overdensity is small at early times ($\delta\propto
t^{2/3}$) and the mass density can be treated as uniform to first
order.  We also assume that the gas and the dark matter are uniformly
mixed.  Because inhomogeneities in the density field are sufficient to
provide tidal torques (Peebles 1969), we do not require that large
numbers of galaxies must have condensed before tidal torquing can be
effective.

(2) As time passes, both the dark matter and baryonic matter begin to
collapse in the overdense region.  Initially, when the densities are
low, the baryons cannot cool, and the pressureless collapse proceeds
identically for both components.

(3) Because the dark matter cannot dissipate energy, its collapse
halts when the system virializes.  Numerical simulations suggest that
the resulting halo has a generic density profile (Warren et al.\ 1992;
Dubinski and Carlberg 1993; Navarro et al.\ 1996): inside the core
radius, the density profile rises roughly as $r^{-1}$, while outside
the core radius, the density profile falls as $r^{-\gamma }$, with
$\gamma $ between 3 and 4. We will approximate the halo with a
Hernquist density profile, whose asymptotic radial profile scales as
$r^{-4}$ (see eqn \ref{hernquist} below).  The size of the core radius
depends only on the mass of the halo and the mean density of the
universe.  For a flat universe (which we will assume for simplicity
for the duration of this paper), the growth of the halo is self-similar,
and the only change in the halo with time is the increase in the
core radius due to its increasing mass.

(4) At some point during the joint collapse of the dark matter and
baryons, the density over some region becomes high enough that the
baryons begin to cool and decouple from the dark matter.  As they
cool, their collapse accelerates, and the baryons begin to concentrate
within the dark matter.  This condensation progressively
increases the baryonic fraction within the inner parts of the
halo. Because the baryons have non-zero angular momentum they cannot
collapse all the way to the center, and instead settle into a rapidly
rotating disk.  The final mass distribution in the disk is determined
both by the initial distribution of specific angular momentum and by
the final rotation curve of the collapsed disk+halo system.

(5) After the onset of cooling, the collapsing baryons further
condense the dark matter halo by increasing the mass density in the
inner parts of the halo, where the baryonic mass fraction has
increased the most.  This modifies the Hernquist potential which the
halo had at the time when cooling began. The condensation of the halo
is assumed to proceed roughly adiabatically.

The calculation outlined above, which we perform explicitly in the
following section, expands upon the formalisms of Fall \& Efstathiou
(1980), Gunn (1982), and van der Kruit (1987) by dropping the
assumption of a static halo and allowing the disk halo to be pulled in
by the collapsing baryons, while retaining the initial condition of a
uniformly rotating protogalaxy and the assumption that the collapse
time is much shorter than the angular momentum transport time.

An essential ingredient in our calculations is the assumption that
angular momentum transport is negligible during the formation of disk
galaxies.  We have several reasons to believe that angular momentum
transport is generally small, whether the formation is smooth, as we
have assumed, or hierarchical, taking place through merging of low
mass subunits.  As we discuss at length in \S\ref{surfbriden}, the
observed angular momenta of disks is comparable to what is expected
for galaxy halos, suggesting that there has not been a dramatic drop
in the angular momentum of the gas during formation.  Furthermore, in
numerical simulations within which the gas fails to conserve angular
momentum during galaxy formation (e.g. Navarro \& Steinmetz 1996), the
disks which form are too compact and have too little angular momentum.
In other words, angular momentum transfer between the gas and the dark
matter halo leads to the formation of galaxies whose density profiles
differ strongly from those of real galaxies.  This empirical argument
suggests that angular momentum transport between the gas and the halo
is generally not important for the formation of disk galaxies.

Angular momentum transport during collapse is most likely to be
negligible for the formation of low surface density disks, which as we
shall show in \S\ref{surfbriden}, are the dominant form of disk
galaxy; angular momentum transport is likely to be most effective in
baryon-dominated systems such as ellipticals and bulges and least
effective in low surface brightness systems.  A pressure supported
dark halo tends to suppress the formation of both bars (Ostriker and
Peebles 1973) and spiral arms (Athanassoula 1984), two effective means
of transporting angular momentum.  In low surface density, high
angular momentum systems, the dark matter dominates the potential, so
bars and spiral arms are unlikely to form.  On the other hand, in low
angular momentum systems, dissipation concentrates the baryons to high
densities, where they are subject to non-axisymmetric instabilities
that will transport angular momentum.  Thus, our basic assumptions are
most likely to be valid for LSBs and are almost certainly not correct
for ellipticals and the bulges of high surface brightness galaxies.
Finally, we expect very little angular momentum transport within the
disk after its formation.  The angular momentum transport time due to
spiral density waves is quite long, $\sim 5-10$ Gyr (Zhang 1996,
Gnedin et al.\ 1995), and thus in the absence of bar formation or
other large scale instabilities, disks cannot redistribute their
angular momentum efficiently once formed.

The other major assumption about angular momentum is that the baryons
have an initial angular momentum distribution similar to that of a
uniformly rotating sphere.  As in past disk formation models, our
assumption is motivated by the similarity between the observed disk
angular momentum distribution and the angular momentum distribution of
a uniformly rotating sphere (Metsel 1963).  Furthermore, in a simple
picture of tidal torquing, most of the torqueing should occur near
maximum expansion and will be due to an external quadrupole.  This
external torque will produce uniform rotation.  Our assumption is also
testable by comparision to numerical simulations, although current
simulations have not addressed this problem explicitly.
 
Finally, in our model we appear to have assumed that all of the
baryons collapse simultaneously, which is physically unreasonable.
The baryons will begin to collapse beyond the dark matter when they
begin to dissipate energy.  The onset of cooling, however, is a
density dependent phenomena, and the baryons in the densest, inner
region of the halo will begin to collapse well before the outer
region.  However, because we are assuming that the system is
spherically symmetric to first order, at any given time both the
baryons and dark matter outside of the cooling region will be largely
unaffected by how the matter inside the cooling region rearranges
itself.  Thus, while the arguments below are dependent on the initial
and final distribution of the baryons, they are independent of the
details of how the collapse proceeds to its final state, as long as
the process is spherically symmetric and adiabatic.

\subsection{Simple Model}	\label{simplemodel}

In this section we will use the above scenario to calculate the mass
profile of the disk and halo after collapse.  The evolution of the
mass profile can be traced through three distinct stages, each
identified with a distinct length scale: 1) the uniform sphere before
collapse, with radius $a$; 2) the Hernquist profile with core radius
$r_0$, at the moment when the majority of the baryons begin to
decouple from the dark matter; and 3) the final baryonic disk,
characterized by an exponential disk scale length $\alpha$.  First we
will use energy conservation to relate the core radius of the
Hernquist profile halo to the radius of the initial protogalaxy.  Next
we will use adiabatic dragging of the halo and the initial angular
momentum distribution of the disk to solve for the final mass profile
and rotation curve of the resulting galaxy.  Finally, we will find the
scale length, central surface brightness, and circular velocity of the
resulting baryonic disk as a function of baryonic mass fraction,
initial angular momentum, and the total galaxy mass.

The core radius $r_0$ of the Hernquist profile can be easily related
to the initial protogalaxy radius $a$, using energy conservation.
Letting $M_{tot}$ be the total mass within the initial sphere of
radius $a$, and $F$ be the fraction of mass within $a$ which is
baryonic, the potential energy of the dark matter in the initial
sphere is $-3(1-F)GM_{tot}^2/5a$.  This must be equal to the potential
energy of the dark matter when the galaxy has partially collapsed into
a Hernquist mass profile, defined as:

\begin{equation}				\label{hernquist}
M_{halo}(r)=(1-F)M_{\infty}\left(\frac{r/r_0}{1 + r/r_0}\right)^2
\end{equation}

\ni where $M_{halo}\left( r\right) $ is the halo mass within radius
$r$.  Its potential energy within a radius $r$ is

\begin{equation}
\Phi\left(x\equiv1+\frac{r}{r_0}\right) =
	-\frac{(1-F)GM_{\infty}^2}{6r_0} \,
	\left[1-\frac{6x^2 - 8x +3}{x^4}\right].
\end{equation}

Fixing the mass within $r=a$ to be constant,
$M_{\infty}=M_{tot}[(a+r_0)/a]^2$.  Assuming that the energy of the
initial uniform sphere is entirely gravitational, that the energy of
the halo within $a$ is conserved during collapse, and that the energy
of the collapsed Hernquist halo is roughly consistent with
virialization ($E\approx\left|\Phi\right|/2$), one finds that
$a=3.2r_0$.  If virialization is not complete, the ratio of $a/r_0$
will be somewhat smaller, but will not significantly change any of our
results.  The relationship between $a$ and $r_0$, as well as all of
the other relations derived in the section, are rederived in the
Appendix for a shallower halo potential suggested
by Navarro et al. (1996).  With the shallower
potential, $a=7.1r_0$.  For a flat universe, the evolution of a dark
matter halo is a scale free process, thus, the density at the core
radius should be a constant multiple of the mean density of the
universe: $r_0=C^{-1}M_{tot}^{1/3}\rho_0^{-1/3}$, where $C\sim 70$
(Navarro et al.\ 1996) and $\rho_0$ is the mean density of the
universe.  This allows the mass scale of the galaxy to be related to
the initial size of the perturbation.

The next step is to calculate the mass distribution of the final disk,
by considering the angular momentum distribution of the baryons and
the adiabatic dragging of the dark matter by the collapsing disk.  We
have assumed that the baryons are evenly mixed with the dark matter
before the onset of cooling, and that they have the distribution of
specific angular momentum characteristic of disks today, which is well
represented by a sphere in solid body rotation (Mestel 1963):

\begin{equation}			\label{mgas}
M_{gas}(<j)=FM_{tot}\left[ 1-\left( 1-j/j_{\max }\right) ^{3/2}\right],
\end{equation}

\ni where $M_{gas}(<j)$ is the baryonic mass of the galaxy which has
specific angular momentum less than $j$, and where $j_{max}=\Omega
a^2$ for a sphere in solid body rotation with angular rotation
velocity $\Omega$.  The maximum angular momentum can also be
characterized by the spin angular momentum parameter $\lambda\equiv
J_{tot}|E|^{1/2}G^{-1}M^{-5/2}$ (Peebles 1969); for our initial conditions
(i.e. an unvirialized sphere in solid body rotation)
$J_{tot}=2M_{tot}j_{max}/5$ and $E=-3GM_{tot}^2/5a$, and therefore
$j_{max}=\frac{5}{2}\left(\frac{5}{3}\right)^{1/2} \lambda
\sqrt{GM_{tot}a}$.

We can now solve for the final mass distribution of the disk+halo
system.  To characterize the radial distribution of halo mass, we
define $m_h(r)\equiv M_{halo}(r)/(1-F)M_{tot}$ to be the fraction of
the halo mass enclosed within a radius $r$.  We will use $m_h$ as a
radial parameter for the rest of this calculation, as well as a
similar parameter related to the enclosed disk mass, $m_d(r)\equiv
M_{disk}(r)/FM_{tot}$.  (We will henceforth drop the explicit
dependence of $m_h$ and $m_d$ upon $r$ for notational convenience.)
Note that the radial dependence of these parameters, $r(m_h)$ (or
$r(m_d)$), will change during collapse, and thus while $m_h$ always
defines a surface bounding a certain fraction of the halo mass, the
radius of that surface is not constant.  After collapse, we will let
$M(m_h)$ ($=M_{tot}\left[(1-F)m_h + F m_d(m_h)\right]$) be the total mass
enclosed within $r(m_h)$.

Assuming that conservation of angular momentum is responsible for
finally halting the collapse of the disk, the specific angular momentum
of gas which halts at radius $r(m_d)$ is $j(m_d)=\sqrt{GM(m_d)r(m_d)}$.
All gas which has specific angular momentum less than $j(m_d)$ will
wind up interior to $r(m_d)$, and thus, using eq. \ref{mgas},

\begin{eqnarray}			
m_d &=& \frac{M_{gas}(<j(m_d))}{FM_{tot}} \\
 &=& 1-\left(1- \frac{\sqrt{GM(m_d)r(m_d)}}{j_{max}}\right)^{3/2}. \label{md}
\end{eqnarray}

Now, to calculate the final mass profile of the halo, the adiabatic
invariance of the angular action $I_\theta$ ($\equiv\int {\bf
v}_\theta \cdot r\,d{\bf \theta}$) must be used (see Binney \&
Tremaine 1987, Chapter 3).  Because the baryons collapse by a large
factor, the time scale for their collapse is much larger than the
orbital time of the dark matter at a given radius; therefore, the
collapse is roughly adiabatic, and the orbits of dark matter particles
will adjust to the changing potential such that $I_\theta^2 = GM(r)r$
is constant (Blumenthal et al.\ 1985, Flores et al.\ 1994).  Using the
inverted form of the Hernquist potential for the initial conditions,
$r_i(m_h)=r_0 m_h^{1/2}/(1+ r_0/a -m_h^{1/2})$, after collapse the
angular action is

\begin{eqnarray}			\label{action}
GM(m_h)r(m_h) &=& G m_h M_{tot} r_i(m_h) \\
	&=& G M_{tot} r_0 \frac{m_h^{3/2}}{1 + (r_0/a) - m_h^{1/2}}.
\end{eqnarray}

\ni Because $M(m_h)r(m_h) = M(m_d(m_h))r(m_d(m_h))$, eqn~\ref{action}
may be substituted into eqn~\ref{md} to solve for $m_d(m_h)$, which
then yields

\begin{equation}			\label{mdmh}
m_d(m_h)= 1 - \left(1 - \frac{\xi m_h^{3/4}}{\sqrt{1 + (r_0/a) -m_h^{1/2}}} \right)^{3/2},
\end{equation}

\ni where $\xi\equiv\sqrt{GM_{tot}r_0}/j_{max}$, or
$\xi=0.173/\lambda$.  Note that for small enough values of $\lambda$,
there is some maximum $m_h$ for which eqn. \ref{mdmh} is well defined;
this cutoff corresponds to the maximum extent of the collapsed disk.
For small values of lambda, the disk collapses to well inside the
final extent of the halo, and beyond this extent, $m_d$ should be
fixed at 1.

The angular action may also be used to derive
the final radius enclosing $m_h$ of the halo mass:

\begin{equation}			\label{rfinal}
r(m_h) = \frac{I_\theta(m_h)} {GM(m_h)}
	= r_0 \left(\frac{m_h^{3/2}}{1 + (r_0/a) - m_h^{1/2}} \right)
	   \left(\frac {1} {\left( 1-F\right)m_h+Fm_d(m_h)}\right).
\end{equation}

\ni Coupled with

\begin{equation}			\label{mfinal}
M(m_h) = M_{tot}\left[(1-F)m_h + F m_d(m_h)\right],
\end{equation}

\ni equations \ref{mdmh}, \ref{rfinal}, and \ref{mfinal} are
parametric equations which fully specify the mass profile of the the
collapsed galaxy.  The final mass profile of the disk and halo
therefore depends only on the baryonic mass fraction $F$, the density
of the universe $\rho_0$, and the mass $M_{tot}$ and spin angular
momentum $\lambda$ of the protogalaxy.  Note that the only place where
our choice of initial conditions (i.e. the uniform sphere) plays a
significant role is in determining the numerical factor relating $\xi$
to $\lambda$.

The above equations can be solved to give both the surface
density and the rotation curve for the final disk:

\begin{equation}			\label{sig_r_mh}
\Sigma(r(m_h)) = \frac{FM_{tot}}{2\pi r(m_h)} \, \frac{dm_d}{dm_h}
			\, \frac{dm_h}{dr}
\end{equation}

\ni and $v_c^2(r(m_h)) = GM(m_h)/r(m_h)$.

As we will show in the following sections, for a $10^{12}M_\odot$ halo
with $F = 0.05$ and $\lambda = 0.06$, the above model implies $r_0 =
35 kpc$, $L = 1.7\times10^{10} L_\odot$, $v_0 = 250$ km/s, $\alpha =
3.4$ kpc, and $\mu_0 =21.2\surfb$, all reasonable parameters for a
massive disk galaxy (assuming a mass-to-light ratio of $3\msun/\lsun$
for the baryons in the disk).

\subsubsection{Surface Density}			\label{surfdensect}

The calculations above solve for the mass of both the dark matter and
the baryons as a function of radius, and thus they can be used to
calculate the surface density profile of the baryonic disk which
results from the collapse.  Figures \ref{surfbrightfig}a and
\ref{surfbrightfig}b show the disk surface density profile which
results from eqn.\ \ref{sig_r_mh} for different values of the
dimensionless spin angular momentum $\lambda$ and the galaxy mass
$M_{tot}$.  All of the disk profiles produced by this gravitationally
self-consistent model are roughly exponential over many scale lengths,
with a scale length $\alpha$ that depends upon $\lambda$, $M_{tot}$,
$F$, and $r_0$.  After fitting the disk profile between 0.5 and 2.5
scale lengths for a wide range of parameters, we derive this
approximate fitting formula for the scale length:

\begin{equation}			\label{alphaeqn}
\alpha(M_{tot}, \lambda, F) \simeq 2.55 r_0(M_{tot})
	 \left[\frac{1}{2+\lambda^{-1}+(0.084/\lambda)^2}\right]^{1+3F}
		\left[16F^2+8F+1+(0.002/F)\right].
\end{equation}

\ni which is good to $\pm5$\% over $0.03<\lambda<0.18$, $0.02<F<0.15$,
and $1\kpc<r_0<120\kpc$; for larger values of $\lambda$, the formula
overestimates the scale length by $\sim10\%$ at $\lambda=0.2$ and
$\sim30\%$ at $\lambda=0.4$.  There is an apparent cut-off in the
disks at large radii in Figure \ref{surfbrightfig} which is due to
using a sharp edged sphere in the initial conditions

Using the fact that the total disk mass is the baryonic fraction $F$ times
the total galaxy mass $M_{tot}$, the approximate central surface brightness is

\begin{eqnarray}					
\Sigma_{0}(M_{tot},\lambda,F)&=&\frac{FM_{tot}}
		   {2\pi\alpha(M_{tot},\lambda,F)^2\Upsilon} \label{sigma} \\
	   &=& \left[\frac{\rho_0^{2/3}C^2}
	            {13\pi\Upsilon}\right] 
		\left[\frac{F}{(1+8F+16F^2+\frac{0.002}{F})^2}\right]\,
 		M_{tot}^{1/3}\,
		\left[2 + \lambda^{-1} +
			\left(\frac{0.083}{\lambda}\right)^2\right]^{2+6F}
\label{surfbrighteqn}
\end{eqnarray}

\ni where $ \Upsilon =(M_{disk}/L_{disk})$; note that $\Upsilon$ is
not a dynamical mass-to-light ratio, but a measure of the efficiency
of turning gas into starlight over the lifetime of the disk.

To first order, $\Sigma_0 \propto
F\,M_{tot}^{1/3}\,\lambda^{-(2+6F)}$, suggesting that low mass and/or
high angular momentum protogalaxies naturally form low surface
brightness galaxies.  Physically, because of angular momentum
conservation, a more rapidly rotating protogalaxy will be unable to
collapse as far inwards as a less rapidly rotating one.  Thus, the
galaxy with a larger specific angular momentum will spread its
baryonic mass over a larger disk, leading to lower mass surface
densities (as seen in eqns. \ref{alphaeqn} and \ref{surfbrighteqn}).
There is observational support for this picture from Zwaan et al.\
(1995) who find that LSBs have systematically larger scale lengths at
a given circular velocity than their high surface brightness
counterparts; given that the specific angular momentum must scale like
$\alpha V_c(\infty)$ to first order, LSBs must have higher angular
momenta than normal galaxies of the same mass.  Most importantly,
Equation \ref{surfbrighteqn} also suggests that {\it any} spread in
either the masses or the angular momenta of protogalaxies translates
into a spread in the resulting distribution of disk surface densities.
We will explore this effect in \S\ref{surfbriden}.

One feature of the models is the presence of a central cusp in the
surface density distribution (Figure \ref{surfbrightfig}).  The mass in
the cusp is only a small fraction of the total disk mass ($\sim6$\%),
and is largely independent of the choice of the initial mass profile
of the halo (see Appendix A).  The cusp is a consequence
of our assumption that the angular momentum distribution is
well-modelled by a uniformly rotating sphere (equation \ref{mgas}).
The center of the disk contains the very low angular momentum material
from the initial protogalaxy, and for a uniformly rotating sphere,
$M(<j)\propto j$ for $j<<j_{max}$, implying $\Sigma\propto r^{-1}$ in
the central regions where the halo contributes very little to the
total mass.  In contrast, the surface brightness distribution of a
pure exponential disk is constant for very small radii, and thus the
surface brightness profiles that result from our models must
necessarily be steeper than exponential in the center.  Because
the cusp is an artifact of our simple initial conditions, we do not
ascribe any profound importance to the bulge-like centers of the
model's light profiles.  Any variation in the shape of the initial
perturbation, the alignment between the angular momentum and the
principle axes of the perturbation, or the degree of angular momentum
conservation will lead to variations in the size and shape of
the cusp, and thus suggest natural mechanisms for producing variations
in the bulge-to-disk ratio.

\vfill
\clearpage
\begin{figure}[ht]
\centerline{ \psfig{figure=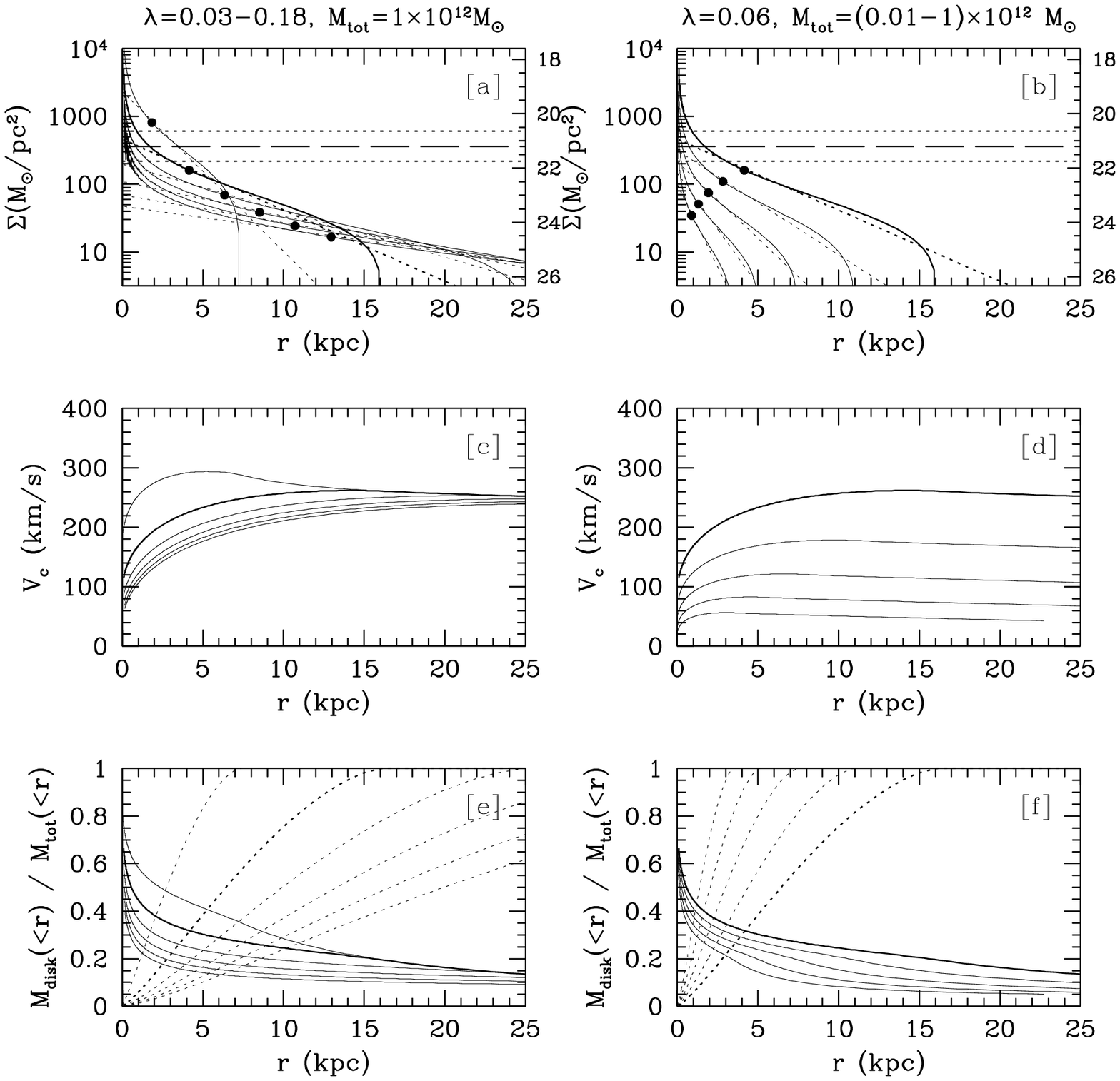} }
\begin{flushright}{\bigskip\cap Figure \ref{surfbrightfig}$\Longrightarrow$}\end{flushright}
\end{figure}
\vfill
\clearpage
\figcaption{\label{surfbrightfig}}
{\small
\ni Resulting surface density profiles [a,b], rotation
curves [c,d], and disk mass profiles [e,f] for disk galaxy models with
various angular momentum (left column; $\lambda=0.03-0.18$, linearly
spaced, $M_{tot}=10^{12}\msun$) and mass (right column;
$M_{tot}=10^{10}-10^{12}\msun$, logarithmically spaced,
$\lambda=0.06$).  In all panels, the heavy line is a fiducial model
with $\lambda=0.06$ and $M_{tot}=10^{12}\msun$; we have assumed
$F=0.05$ and $H_0=50\hnot$ for all the models.  The top row [a-b]
shows the surface density as a function of radius, with the right hand
axis giving the apparent surface brightness $\mu_0$ assuming a disk
mass-to-light ratio of $\Upsilon=3\msun/L_\odot$.  The horizontal
dotted lines bracket the Freeman (1970) surface brightness.  The solid
lines are the models, the dotted diagonal lines are exponential fits
to the surface brightness distribution, and the solid dots are at one
exponential scale length ($r=\alpha$).  The models produce roughly
exponential profiles over several scale lengths.  Panel [a] shows how
increasing the angular momentum of the protogalaxy decreases the
extrapolated central surface brightness and increases the exponential
scale length of the resulting disk, while panel [b] shows how
increasing the mass of the protogalaxy increases both the central
surface brightness and the scale length of the resulting disk.
The middle row [c-d] shows the rotation curves which result from the
models.  Panel [c] shows that changing the angular momentum changes
the shape of the rotation curve, such that high angular momentum
models rise more slowly, but keeps the same asymptotic circular
velocity.  Panel [d] shows that decreasing the mass decreases the
asymptotic circular velocity; the shape of the rotation curve does not
change however, and simply rescales with the exponential scale length
and mass.  The bottom row [e-f] shows the baryonic-to-total mass
fraction within a given radius (solid lines), as well as the fraction
of the baryonic mass contained within a given radius (dotted lines).
These plots show the degree to which the dynamics of the galaxy are
strongly affected by the baryons.  Panel [e] shows that in low angular
momentum models ($\lambda\le0.04$), almost all of the baryonic mass
lies within a radius where baryons make up more than 40\% of the total
mass, whereas in high angular momentum models ($\lambda\ge0.09$),
almost none of the baryonic mass falls in a region where baryons make
up more than 40\% of the total mass.  The models also suggest that in
the high angular momentum, low central surface brightness models, the
halo dominates the dynamics over almost the entire disk.  Panel [f]
shows that for changing mass, the baryonic mass fraction is
self-similar, scaling with the exponential scale length.
}
\newpage


\subsubsection{Rotation Curves}			\label{rotcurvesect}

Equations \ref{rfinal} and \ref{mfinal} can readily be used to
calculate the circular velocity as a function of radius.  Figures
\ref{surfbrightfig}c and \ref{surfbrightfig}d show the resulting
rotation curves for the same disk plus halo models shown in Figures
\ref{surfbrightfig}a and \ref{surfbrightfig}b respectively, for
different values of the spin angular momentum $\lambda$ and the total
mass $M_{tot}$. For simplicity, we have included only the monopole
term from the disk and not thus included the effects of disk
flattening.  The rotation curves confirm Blumenthal et al.'s (1985)
and Flores et al.'s (1995) conclusion that adiabatic contraction of
the halo leads naturally to asymptotically flat rotation curves, and,
as previous authors have found, that for a wide range of parameters,
the rotation curves of the models look much like those of real galaxies.

Figure \ref{surfbrightfig}[c\&d] demonstrates how changes in mass and
angular momentum affect the shape of the rotation curve $V_c(r)$.
First, the shape of the rotation curve remains constant with changes
in mass, when scaled by the exponential scale length, for fixed spin
angular momentum $\lambda$.  At fixed angular momentum, the baryons
collapse by the same factor within the self-similar dark matter halos,
leading to identical fractions of dark and luminous matter within a
given dimensionless radius $r/r_0$ (or $r/\alpha$), and thus to
self-similar rotation curves.  However, if there is any deviation from
perfect self-similarity in the shape of dark matter halos, then there
will be accomanying deviations in the self-similarity of the rotation
curves.  Violations of our assumption of perfect self-similarity will
lead to either increased scatter in the shape of rotation curves at
fixed angular momentum, or to systematic variations in the shape of
rotation curve with changing mass.

In contrast, changes in the spin angular momentum $\lambda$ lead
directly to variations in the shape of the rotation curve.  Increasing
the spin angular momentum of a galaxy decreases the collapse factor of
the baryons (i.e.\ $r_i/r_f$), leaving them at a larger dimensionless
radius.  Thus, an increase in the spin angular momentum reduces the
fraction of baryonic mass within any given dimensionless radius,
leading to a rotation curve whose shape is dominated more by the dark
matter distribution than the baryonic matter.  There are two strong
observational signals of these high angular momentum rotation curves.
First, high angular momentum galaxies will have a larger dynamical
mass-to-light ratio, due to the increased dominance of dark matter
over baryons at every radius (as can be seen in the mass profiles of
Figure \ref{surfbrightfig}[e\&f]).  Second, the rotation curves will
rise gradually with radius, instead of steeply as is usually observed
for more concentrated low angular momentum disks.

As shown in \S\ref{surfdensect}, the models suggest that such high
angular momentum systems should to be found preferentially among the
population of low surface brightness galaxies (eqn
\ref{surfbrighteqn}).  Indeed, the rotation curves of LSBs do reveal
the same signatures as the high-angular momentum models.  Edge-on
(Goad \& Roberts 1981) and face-on LSBs (de Blok et al 1996) show a
systematically slower rise in their rotation curves than do their high
surface brightness counterparts.  De Blok et al (1996) also find that
the dynamical mass-to-light ratio increases systematically with
surface brightness, in concordance with the model.

The rotation curves in Figure \ref{surfbrightfig}[c\&d] and the mass
profiles in Figure \ref{surfbrightfig}[e\&f], suggest that low surface
brightness disks are extremely good tracers of the properties of dark
matter halos.  First, because low surface density disks contribute
less to the dynamics of the galaxy within any given radius, the
resulting rotation curve is much closer to what would be expected for
a massless disk.  Second, LSB disks will tend to reflect any
deviations from axisymmetry within the halo.  Because the disks do not
have enough mass to round out the potential at a given radius, the
disks should be more asymmetric in general; however, if LSBs are high
angular momentum systems, they may preferentially form in halos with
large quadrapole moments, which will lead to a further increase in the
apparent disk asymmetry.  Finally, LSBs have larger scale lengths than
high surface brightness galaxies of the same mass.  This allows the
halo to be traced to a much larger radius, allowing the mass profile
of the halo to be measured at extremely large distances.

\begin{figure}
\centerline{ \psfig{figure=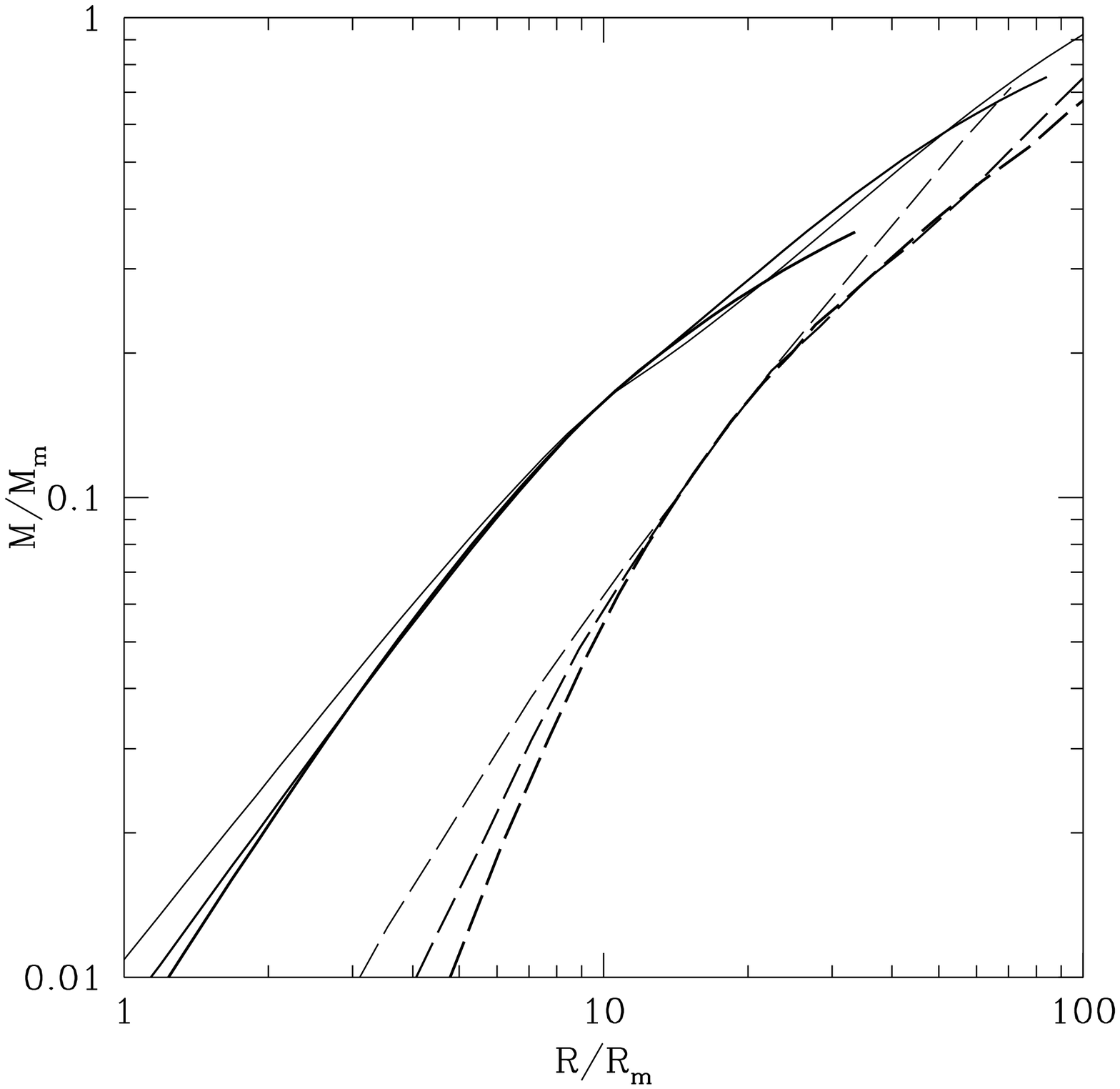,height=5in} }
\figcaption{Comparison between the shapes of the rotation curves resulting
from models with $\lambda=0.03$ (light solid), $\lambda=0.05$ (medium
solid), and $\lambda=0.03$ (heavy solid), and the Burstein \& Rubin
(1985) ``mass types'' I (light dashed), II (medium dashed), and III
(heavy dashed).  The curves have been rescaled in radius and mass to
match at their points of maximum curvature, as in Burstein \& Rubin
(1985).  Changes in angular momentum are clearly capable of producing
variations in shape comparable to those found observationally by
Burstein \& Rubin.  The models rise less steeply in the center and
more slowly in the outer regions than do the Burstein \& Rubin mass
types.  This probably reflects a difference between reality and our
assumed halo profile or angular momentum distribution, as well as
possibly our neglect of bulges.
\label{bursteinfig}}
\end{figure}

We may make a detailed comparison between the shapes of observed
rotation curves and the rotation curves produced by the models.
Figure \ref{bursteinfig} shows the $\log{R}$-$\log{M(R)}$ relations
for the Burstein \& Rubin (1985) ``mass types'' (derived from observed
rotation curves) and the identical relation for models spanning a
moderate range of spin angular momenta.  (All of the curves have been
rescaled by an arbitrary factor $R_m$ and $M_m$ to match at their
point of maximum inflection (as in Burstein \& Rubin 1985), and the
model curves have been shifted upwards from the observations to make
the comparison easier.)  Figure \ref{bursteinfig} shows that, one,
both the models and the observations are remarkably similar, with the
rotation curves being well approximated a broken power-law, and two,
that it is the faster rising rotation curves which have less
curvature.  In the models, these changes in shape are produced by
reasonable variations in the spin angular momentum.  Furthermore,
because changes in mass do {\it not} produce changes in the shape of
the rotation curve, there should not be any strong correlation between
galaxy mass and the distribution of Burstein \& Rubin ``mass types'',
unless the dark matter halos are not self-similar with mass, as we
have assumed.  There is a tendency for the model mass distributions to
be slightly shallower at small radii than the observed distributions,
which probably reflects either a deviation from the assumed initial
Hernquist halo mass profile or in the initial distribution of angular
momentum.  Given the simplicity of the initial assumptions, this is
less notable than the general degree of agreement.

\vfill
\clearpage
\vfill
\clearpage

Recent work by Persic, Salucci, \& Stel (1996; see also Persic \&
Salucci 1996) has parameterized the shape of galaxy rotation curves as
a function of mass to produce a ``Universal Rotation Curve'' (URC).
However, as our modelling shows, such attempts may perhaps reveal more
about the selection criteria of the sample of galaxies used to
generate the URC than about the rotation curves themselves.  First,
because mass and angular momentum are fundamentally linked with
surface brightness and exponential scale length, any selection bias
which affects the distribution of surface brightnesses and scale
lengths present in a galaxy sample will critically affect any URC
derived from the sample; any selection effect which changes the
distribution of surface brightnesses and scale lengths in a sample can
change the apparent distribution of disk angular momenta, and thus the
distribution of rotation curve shapes.  For example, intrinsically
large scale lengths will be overrepresented in angular diameter
limited surveys of field galaxies, while an angular diameter limited
surveys of cluster galaxies will have a larger fraction of galaxies
with small scale lengths. Thus the two surveys will have different
distributions of galaxy masses and angular momenta, giving different
apparent relations between mass and rotation curve shape.
Furthermore, there are unavoidable biases against finding low surface
brightness galaxies, which will tend to reduce the contribution of low
mass or high angular momentum galaxies to the derived URC.  To avoid
drawing any erroneous conclusions, any parameterization of rotation curve
shapes must include a detailed discussion of the selection criteria
that went into choosing the galaxies whose rotation curves were
measured.

\subsubsection{Tully-Fisher Relation}		\label{TFsec}

In addition to understanding the shape of galaxy rotation curves,
our models may be used to understand the behavior of the Tully-Fisher
relation as a function of surface brightness.  Towards that end,
fitting the circular velocity at three disk scale lengths,
we find

\begin{equation}				\label{Vc}
V_c(r=3\alpha;r_0,\lambda,F) = 8\kmsec\left[\frac{r_0(M_{tot})}{\kpc}\right]
			\left[a(\lambda)F + b(\lambda)\right]
\end{equation}

\ni where

\begin{equation}
a(\lambda) = 1 + 35\left[\frac{(0.015/\lambda)^2}{1+(0.015/\lambda)^2}\right]
\end{equation}

\ni and

\begin{equation}
b(\lambda) = 0.22 + 0.68\left[\frac{(\lambda/0.015)}{1+(\lambda/0.015)}\right],
\end{equation}

\ni which is accurate to $\pm5$\% for $0.01<\lambda<0.14$,
$0.01<F<0.15$, and $1\kpc<r_0<120\kpc$.  The lambda dependence (i.e.\
the term in brackets) in equation \ref{Vc} contributes only a
$\pm20$\% variation in $V_c$ for $\lambda>0.025$.  For very small
angular momentum models, however, the concentration of the disk is
substantial enough that the mass of the compact disk contributes
significantly to the rotation curve, giving a factor of two variation
in $V_c$ for $\lambda<0.025$.  However, these systems are the least
likely to survive as pure disk galaxies; these low angular momentum
disks are disk dominated in their centers and are likely to be
unstable to bar formation.  The bar would transfer angular momentum
outwards to the halo (Weinberg and Hernquist 1992), which
redistributes angular momentum within the galaxy and alters its
structure. The low angular momentum systems are also plausible
candidates for elliptical galaxies.

\begin{figure}[hb]
\centerline{ \psfig{figure=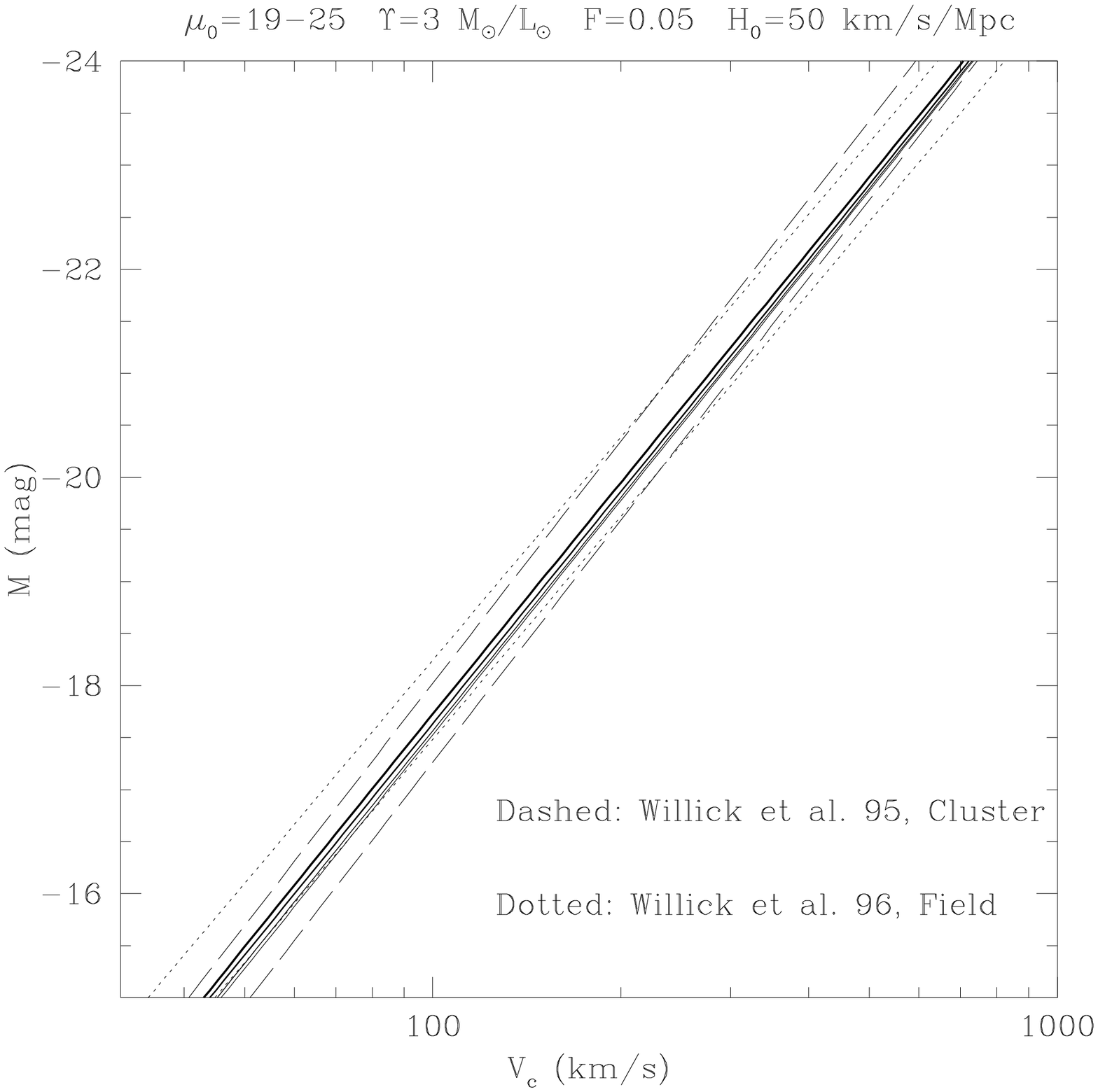,height=3in} }
\figcaption{\label{T-Ffig}}
{\footnotesize
\addtolength{\textwidth}{-1in}
\ni Comparison between the Tully-Fisher relation predicted by
equation \ref{Vc} for $\mu_0=19-25\surfb$ (solid lines, lighter
corresponding to fainter surface brightness) and the observed
Tully-Fisher relation ($\pm 1\sigma$) for clusters (dashed line;
Willick et al.\ 95) and the field (dotted line; Willick et al.\ 96).
Note that, for the assumption of a constant mass-to-light ratio for
galactic disks, there should be no shift in the mean Tully-Fisher
relation as a function of surface brightness.  We have assumed
$\Upsilon=3\msun/L_\odot$, $F=0.05$, and $H_0=50\hnot$ for the models.
}

\end{figure}
\clearpage

Equation \ref{Vc} above can be used to understand the Tully-Fisher
relationship (Tully \& Fisher 1977; see Strauss \& Willick 1995 for a
review) for disk galaxies, for {\it all} surface brightnesses.
Because the scale length of the disk is proportional to $r_0$ (eqn.\
\ref{alphaeqn}), and the circular velocity at fixed scale length is
proportional to $\sqrt{M(r)/r_0}$, giving $L\propto V_c^3$, with a
numerical constant that depends upon $F$, $\Upsilon$, and $\lambda$.
This slope is consistent with the observed infrared Tully-Fisher
relation (Strauss and Willick 1995).  Furthermore, because of the weak
dependence of $V_c$ upon $\lambda$, equation \ref{Vc} predicts very
little offset in the Tully-Fisher relation with changes in $\lambda$,
and thus predicts that the Tully-Fisher relations for both high and
low surface brightness galaxies should be nearly coincident.  We show
this in Figure \ref{T-Ffig}, where we plot the relationship between
circular velocity and absolute magnitude predicted by equation
\ref{Vc} for a wide range of disk surface brightnesses.  Superimposed
are the $\pm1\sigma$ limits for the Tully-Fisher relation derived by
Willick et al (1995, 1996) for normal cluster and field galaxies.  The
predicted Tully-Fisher relations fall well within the observed scatter
over a factor of 100 in surface brightness and a factor of 1000 in
mass.

The independence of the predicted Tully-Fisher relation and disk
surface density is not surprising given that the relation primarily
results from the properties of the dark matter halo, if one reasonably
assumes that the luminosity of a galaxy is proportional to the total
mass, and that the measured line width reflects the asymptotic
circular velocity at large radii, where the mass is dominated by the
halo; the relationship between luminosity and line width should not
depend strongly on how the baryons are distributed within the halo.
However, the predicted relations do assume a single disk mass-to-light
ratio $\Upsilon$ for all masses and all disk surface densities; any
variation in $\Upsilon$ should translate into increased scatter in the
relation, and any systematic variation in $\Upsilon$ with the total
mass or surface density of the disk should translate into changes in
the Tully-Fisher slope.  Therefore, the remarkable agreement between
the predicted Tully-Fisher relation in Figure \ref{T-Ffig} and the
measured one implies that there is very little evidence for a
systematic variation of the disk mass-to-light ratio with mass.  On
the other hand, while Sprayberry et al.\ (1995) and Zwaan et al.\
(1995) both find that LSB galaxies fall on the same Tully-Fisher
relation defined by normal galaxies, they also find increased scatter
in the relationship, roughly by factors of two.  This suggests that
there is more variation in the disk mass-to-light ratio for low
surface density disks than for high surface density disks; given the
more tenuous nature of these disks, we do not find this conclusion to
be unreasonable.  However, given that the mean Tully-Fisher relation is
indistinguishable for both high and low surface brightness galaxies,
the mean disk mass-to-light ratio must be similar across the range of
observed surface brightnesses.

\subsection{Pressure Support}			\label{jeans}

In the preceding calculation of the galaxy formation, we assumed that
the baryons will collapse along with the dark matter halos.  However
the baryonic gas experiences pressure forces in addition to
gravitational forces.  For small masses, internal pressure may support
the gas against collapse, in spite of the inward gravitational pull of
the collapsed dark matter halo (Jeans 1929).  Because of the link
between low masses and low surface brightnesses (eqn
\ref{surfbrighteqn}), there is some limiting surface density for which
we expect our assumption of complete baryonic collapse to become
invalid.

To contrain the regime of galaxy properties for which pressure support
can affect the formation of a disk, we consider the simpler question:
what is the range of halo properties for which the gas is bound to the
dark matter halo?  The gas will be bound if the gravitational binding
energy is greater than its thermal energy.  Assuming that the gas is
fully ionized, it will be gravitationally bound to the dark halo if

\begin{equation}					\label{thermgrav}
\frac{m_p V_c^2}{2} > k T,
\end{equation}

\ni where $V_c$ is the circular velocity of the dark matter
halo, $m_p$ is the proton mass, and $T$ is the temperature
of the gas.  Applying this relation at the halo half mass radius,
where $v_c = \sqrt{GM/r_0} = \sqrt{GC} M^{1/3}\rho_0^{-1/6}$
implies a minimum halo mass for collapsed objects:

\begin{equation}					\label{Mmineqn}
M_{\min} = \rho_0 \left(\frac{k T} {G \rho_0 m_p C}\right)^{3/2} = 4 \times 10^7 M_\odot T_4^{3/2}
\end{equation}

\ni where $T_4 = T/10^4 K$.  Thus, the Jeans criterion limits the
formation of low mass galaxies, rather than low surface brightness
galaxies, and we can treat the above calculation as valid for
all protogalaxies with $M_{tot}>M_{min}$.

\section{The Number Density of Disk Galaxies}		\label{surfbriden}

The equations of \S\ref{surfdensect} provide a means to relate the
primordial quantities of mass and angular momentum to the observable
properties of disk galaxies, namely central surface brightness and
scale length.  Theoretical models and numerical simulations provide a
way to predict the distribution of mass and angular momentum for
galaxy halos, and thus we have the means to predict the observed
distribution of disk central surface brightnesses ($\mu_0$) and scale
lengths ($\alpha$).  Furthermore, because most selection effects in
galaxy catalogs can be modelled as detection efficiencies as a
function of surface brightness and/or scale length, the resulting
joint distribution of $\mu_0$ and $\alpha$ can be convolved with a
model of a survey's detection efficiency to predict the observed
distribution of either $\mu_0$ and $\alpha$, or equivalently, surface
brightness and luminosity.

To begin, we must first assume an intrinsic distribution
for spin angular momenta, $p(\lambda)$.  This problem has been
approached both analytically and numerically by a number of groups
(Barnes \& Efstathiou 1987, Ryden 1988, Warren et al. 1992, Eisenstein
\& Loeb 1995, Catelan \& Theuns 1996a, Catelan \& Theuns 1996b), all
of which find the same general properties.  To first order, the
distribution of spin angular momentum of collapsed dark matter halos
is well approximated by a log normal distribution:

\begin{equation}		\label{spindisteqn}
p(\lambda )=\frac {1}{\sigma_\lambda\sqrt{2\pi }} \exp \left( -\frac{\ln
^2(\lambda /\langle \lambda \rangle )}{2\sigma_\lambda ^2}\right)
\frac{d\lambda }{\lambda}.
\end{equation}

Most work finds that this distribution peaks around $\langle \lambda
\rangle \approx 0.05$, with a width in the log of $\sigma_\lambda
\simeq 1$.  For the purposes of this paper, we adopt the specific
values of $\sigma_\lambda \simeq 0.7$, and $\langle \lambda
\rangle \simeq 0.06$, which provide a good fit to the results of
Warren et al. (1992).  

For simplicity, we will
assume that $p(\lambda)$ is identical for all sizes of the initial
density perturbation (i.e. angular momentum is independent of mass).
Faber (1982) has argued that the spin angular momentum of a
protogalaxy should be independent of the amplitude of the initial
overdensity, based upon scaling laws.  However, there is some evidence
from analytic calculations of $p(\lambda)$ in the peak-height
formalism which suggests that the mean spin angular momentum, $\langle
\lambda \rangle$, is a function of the height of the peak in the
initial density field which forms a given protogalaxy (Eisenstein \&
Loeb 1995, Catelan \& Theuns 1996a):  The trend is
such that smaller initial peaks, which are likely to be more
asymmetric, are more subject to torques and have larger mean angular
momenta; this would make it even more likely to find that low mass
galaxies are also low surface brightness galaxies, and could
potentially lead to systematic variations in galaxy-galaxy correlation
properties as a function of surface brightness.  However, given that
the shift in angular momentum distribution with peak height has yet to
be verified numerically, we will retain the simple assumption of the
independence of angular momentum and peak height.  

Another of our assumptions is that the angular momentum of the gas is
conserved during collapse, independent of the parameters of the
collapse.  This assumption should be robust when a collapse can be
approximated as smooth; for example, in a collapse where there are no
major mergers, or where the gas in merging subclumps is extended and hot
enough that there is little angular momentum transfer between the gas
clumps and the dark matter halos.  However, the degree to which angular
momentum conservation holds may well depend on the detailed merger
history of a given galaxy, which in turn could depend systematically on
mass, angular momentum, and peak height.

While there are theoretical and numerical tools for approximating the
merger history of galaxy halos of a given size (e.g. Bower 1991, Bond
1991, Lacey \& Cole 1993, 1994), there is not a consistent picture of
the behavior of gas in these merging systems which leads to the
observed galaxy structure.  For example, depending on the assumptions
about the ultraviolet background radiation and star formation
feedback, galaxies in hierarchical formation scenarios can have from
30\% to 90\% of their gas cool (Navarro \& White 1993, Evrard,
Summers, \& Davis 1994, Thoul \& Weinberg 1996, Navarro \& Steinmetz
1996).  A consistent feature is that these simulations produce disks
in which the gas is much more centrally concentrated than in disks of
real spiral galaxies.  The spin parameters of the simulated disks show
a range of values, generally below those observed, and indicate some
angular momentum transport to the dark matter.  However, given the
developing nature of the field and, more importantly, the failure to
reproduce the observed gas profiles, it would imprudent to assume
angular momentum transport plays a strong role in galaxy formation and
we have no firm basis for considering its variation with the
properties of galaxy halos.

In principle, any modification can be worked into equation
\ref{spindisteqn}, but, for now, we take the properties of the dark
matter halos as representative of the properties of the resulting disks.
This simplification may eventually prove to be a poor assumption for
some individual galaxies, but it would be incorrect in general only if
there were such extensive, violent merging as to entirely erase any
memory of the initial angular momentum distribution from the resulting
gas distributions, for the bulk of the galaxy population.  Such a
scenario seems unlikely as it raises serious questions about why spiral
galaxies have angular momenta which are comparable to the predicted
angular momenta of dark matter halos and how one avoids excessive
central gas concentrations in such a model.

In addition to the above distribution of spin angular momenta, to
fully specify the predicted distribution of disk central surface
brightnesses and scale lengths we also need a prediction for the
distribution of galaxy masses.  As a first approximation, we use a
Schechter function with a power law of slope $\alpha$ at the low mass
end, with an exponential cutoff at masses greater than $M_*$, such
that

\begin{equation}
n(M_{tot})=\Phi_*(M_{tot}/M_*)^{\alpha_{lum}} \exp{(-M_{tot}/M_*)}.
\end{equation}

\ni(Press \& Schechter 1974, Schechter 1975).  

With the above distribution of spin angular momenta and galaxy masses,
we may calculate the predicted distribution of disk central
surface brightnesses and scale lengths.  First, 
we may use the distribution of angular momentum in equation
\ref{spindisteqn} and the relationship between surface
brightness and angular momentum in equation \ref{surfbrighteqn} to
calculate the predicted distribution of disk surface brightness
at fixed disk mass:

\begin{equation}			\label{psigMdeqn}
p(\Sigma_0|M_d)=p(\lambda)\left|\frac{d\lambda}{d\Sigma_0}\right|_{M_d},
\end{equation}

\ni which requires the following inversion of equation \ref{surfbrighteqn}:

\begin{equation}
\lambda(\Sigma_0|M_d) =
	\frac{A + \sqrt{0.944A^2 + 0.028A}}{2-4A},
\end{equation}

\ni where

\begin{equation}
A=\left[\frac{\sqrt{M_d/2\pi\Upsilon\Sigma_0}}{2.55r_0(1+8F+16F^2+\frac{0.002}{F})}\right]^\frac{1}{1+3F}.
\end{equation}

\ni Then, with equations \ref{alphaeqn} and \ref{psigMdeqn} and the
distribution of galaxy masses $n(M_d)$, the joint distribution of
central surface brightness and scale length is

\begin{eqnarray}			\label{psigalphaeqn}
n(\alpha,\Sigma_0)&=&n(M_d,\Sigma_0)\left|\frac{dM_d}{d\alpha}\right| \cr
		&=&p(\Sigma_0|M_d)n(M_d(\Sigma_0,\alpha))
			\left|\frac{dM_d}{d\alpha}\right|.
\end{eqnarray}

\ni All of the above equations involving the central surface
brightness $\Sigma_0$ can easily be reformulated in terms of magnitude
per square arcsecond, $\mu_0$, using the conversion
$p(\mu)=p(\Sigma_0)\Sigma_0/(2.5\log{e})$.  

Because the distribution of angular momenta (eqn.\ \ref{spindisteqn})
is broad, spanning more than a factor of 10 in $\lambda$, equations
\ref{surfbrighteqn} and \ref{psigMdeqn} suggest that the distribution
of galaxy surface brightnesses should also be broad\footnote{Note that
van der Kruit (1987) came to a nearly opposite conclusion with a
similar analysis, by assuming that disk galaxies had one unique value
of $\lambda$.  His Equation 15 (an analog to our equation~\ref{sigma})
therefore predicted a single characteristic disk surface density.}.
For example, keeping only the $\lambda^{-2-6F}$ dependence of equation
\ref{surfbrighteqn}, the log-normal distribution of angular momenta
(eqn.\ \ref{spindisteqn}) leads to a Gaussian distribution of central
surface brightness $\mu_0$, with width $\sigma_\mu =
2.5(2+6F)\log{(e)}\sigma_\lambda$, or roughly
$2.5\sigma_\lambda\surfb$.  Given that most numerical studies find
$\sigma_\lambda\sim0.7$, there is immediately reason to expect
galactic disks to form with a very wide distribution of central
surface brightnesses, at every luminosity.

The peak of the (approximately) Gaussian surface brightness
distribution corresponds to a characteristic surface brightness for
each given mass:

\begin{eqnarray}				\label{charmueqn}
\langle\mu_0\rangle &=& 27.07-2.5\lg{(\left[\frac{\rho_0^{2/3}C^2}
	            	{13\pi\Upsilon}\right] 
			\left[\frac{F}{(1+8F+16F^2+\frac{0.002}{F})^2}\right]\,
 			M_{tot}^{1/3}/\langle\lambda\rangle^{2+6F}})\cr
		&=& 22.7 + 2.5\lg{\frac{\Upsilon}{3\msun/\lsun}} - 0.83\lg{\left[\frac{M_d}{10^{10}\msun}\right]},
\end{eqnarray}

\ni in units of $B$ magnitudes per square arcsecond, assuming
$F=0.05$.  The characteristic surface brightness for ``normal'' bright
galaxies is therefore comparable to the canonical Freeman (1970) value
of $21.7\pm0.3\surfb$ for disks.  However, the expected width of
$\pm3.4$ magnitudes suggests that there will be large numbers of
bright galaxies with very faint disk surface brightnesses, down to
$\sim24\surfb$; at lower luminosities, the expected distribution of
disk properties will shift to even lower surface brightnesses.

\vfill
\subsection{The Distribution of Surface Brightness and Scale Length} \label{mualphasec}

The general picture that the above sections describe is one where the
structure of galaxy disks can be directly linked to mass and angular
momentum through equations \ref{alphaeqn} and \ref{surfbrighteqn}.  At
fixed angular momentum, low mass galaxies have faint central surface
brightnesses and small scale lengths.  At a fixed mass, galaxies with
large angular momentum spread their mass over a larger area, leading
to large scale lengths but faint central surface brightnesses.
Therefore, we expect to find galaxies with a wide range of surface
brightnesses and scale lengths, due to variations in mass and angular
momenta among protogalaxies.  We may take the distribution of surface
brightness and scale length predicted by this scenario (eqn.\
\ref{psigalphaeqn}) and compare it with the observed properties of
the population of disk galaxies.

\vfill
\clearpage
\begin{figure}[ht]
\centerline{ \psfig{figure=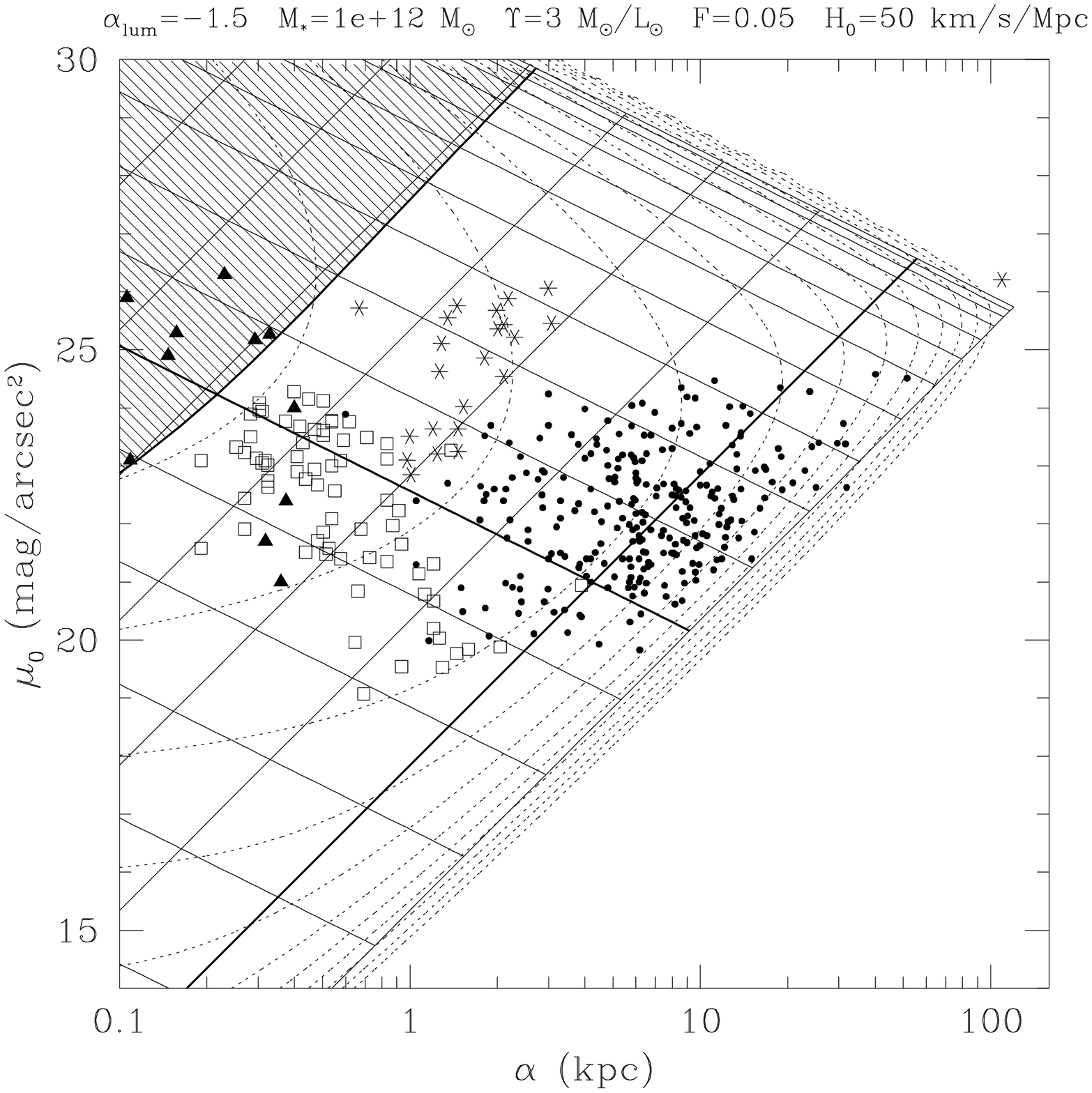} }
\begin{flushright}{\bigskip\cap Figure \ref{alphamufig}$\Longrightarrow$}\end{flushright}
\end{figure}
\vfill
\clearpage
\figcaption{\label{alphamufig}}
{\ni
\small
Observed and predicted distribution of exponential scale lengths
$\alpha$ and central surface brightness $\mu_0$ (in the $B$ band) for
galactic disks.  The solid circles are taken from a variety of sources
(Romanishin et al.\ 1973, Boroson 1981, van der Kruit 1987, Sprayberry
et al.\ 1995, de Blok et al.\ 1995, McGaugh \& Bothun 1994, de Jong \&
van der Kruit 1994, de Jong 1996, Knezek 1993), and consist mostly of
galaxies chosen from the NGC, the UGC, or the POSS-II catalog of LSBs
(Schombert et al.\ 1992), all of which are either explicitly or
implicitly angular diameter limited field surveys; however, because of
how the points were assembled from the literature, the filled circles
cannot be used to determine the relative number densities of galaxies
as a function of $\alpha$ and $\mu_0$, although they do give a good
indication of the range of galaxy properties found in field surveys.
The open squares are from the center of the Virgo cluster (Bingelli et
al.\ 1984), the stars are from a catalog of LSBs in the direction of
the Virgo cluster (Impey et al.\ 1988), including Malin I (upper far
right).  The filled triangles are the local group dwarf spheroidals
compiled by Caldwell et al.\ 1992.  The dotted lines are contours of
constant $n(\lg{\alpha},\mu_0)$ (from equation \ref{psigalphaeqn}),
assuming $\Upsilon=3\msun/L_\odot$, $F=0.05$, and $H_0=50\hnot$ (which
are the same parameters which fit the Tully-Fisher relation in Figure
\ref{T-Ffig}), and assuming $\alpha_{lum}=-1.5$, $M_*=10^{12}\msun$,
$\langle\lambda\rangle=0.06$, and $\sigma_\lambda=0.7$ for the mass
and angular momentum distributions.  The diagonal solid lines of
positive slope are lines of constant mass (separated by factors of 10,
with the heavier line at the characteristic mass $M_*$), and the
diagonal solid lines of negative slope are lines of constant angular
momentum (logarithmically spaced, separated by factors of $10^{0.2}$,
with the heavier line at a characteristic angular momentum
$\lambda>\sim0.06$).  The cross-hatched region in the upper left
corner is the area where gas pressure will prevent galaxies from
collapsing (i.e. $M<M_{min}$; \S\ref{jeans}).  Note that: 1) the
region of the plane occupied by galaxies is strongly dependent on the
type of survey used to find the galaxies; 2) cluster surveys, which
are largely unbiased with regard to physical scale length, are
dominated by galaxies with low surface brightnesses and small scale
lengths, as would be predicted by $n(\lg{\alpha},\mu_0)$; 3) both the
data and the models show the same paucity of physically large, high
surface brightness galaxies, which reflects the exponential mass
cutoff in the mass function of galaxies; and 4) galaxies exist with
the lowest surface brightnesses to which the deepest surveys are
sensitive, suggesting there are likely to be galaxies with even lower
surface brightnesses than have currently been observed.  See
\S\ref{mualphasec} for a fuller discussion.

}
\vfill
\clearpage

In Figure \ref{alphamufig}, the distribution of surface brightness and
scale length predicted by equation \ref{psigalphaeqn},
$n(\lg{\alpha},\mu_0)$, is drawn as dotted contours separated by
factors of 10 in number density.  Superimposed upon these contours are
diagonal lines of constant disk mass (the solid lines of positive
slope) and lines of constant angular momentum (the solid lines of
negative slope).  The distribution plotted in Figure \ref{alphamufig}
shows several observationally testable features.  First, the
distribution is extremely broad in both surface brightness and scale
length, suggesting that disk galaxies should form with an extremely
wide range of properties.  Second, looking at $n(\lg{\alpha},\mu_0)$
along a line of constant mass, the distribution suggests that galaxies
of any mass have a wide range of surface brightnesses.  Thus, low
surface brightness galaxies are not necessarily dwarf galaxies, and
many examples should be found with masses which are comparable to
normal galaxies.  Third, both normal and low surface brightness
galaxies should show the exponential cutoff with mass.  At the cutoff,
low surface brightness galaxies will have systematically larger scale
lengths due to their high angular momentum.  These effects lead to an
apparently larger maximum disk scale lengths with decreasing surface
brightness (i.e. the diaganol edge in Figure \ref{alphamufig} towards
high surface brightness and large scale length).  Fourth, the number
density rises with decreasing scale length and decreasing surface
brightness, suggesting that there is an enormous population of small,
low surface brightness galaxies.  The rise is due to a combination of
the increase in the galaxy mass function with decreasing mass, and of
the correlation between mass and surface brightness and/or scale
length (eqns \ref{sigma} \& \ref{alphaeqn}).  Finally, Figure
\ref{alphamufig} suggests that there should be a correlation
between mass and surface brightness.  The correlation results from eqn
\ref{sigma}, which suggests surface brightness is proportional to
$M^{1/3}$.  However, as can be seen by the contours of
$n(\lg{\alpha},\mu_0)$, the correlation will be broad, due to the
width of the angular momentum distribution.  The correlation may be
the source of the surface brightness - luminosity relation found by
Bingelli et al.\ (1984) in the Virgo cluster, and by Ferguson
\& Sandage in Fornax (1988).

To compare the predicted distribution of $n(\lg{\alpha},\mu_0)$ to
observations of disk galaxies, in Figure \ref{alphamufig} we have also
plotted observed properties of galactic disks for a variety of surveys
with different selection techniques.  Because selection effects are
largely responsible for shaping the apparent distribution, for the
moment we will restrict our comparison to galaxies discovered
exclusively in large photographic angular diamter limited field
surveys (i.e. NGC, UGC (Nilson 1973), POSS-II LSB survey (Schombert et
al.\ 1992), APM LSB survey (Impey et al.\ 1996)), all of which have
similar angular diameter limits and thus similar selection criteria.
These field galaxies are plotted as filled circles in Figure
\ref{alphamufig}, based upon their published surface brightness
photometry.  While little can be said about the relative number
density of the field galaxies on this plot, given that the disk
properties are drawn from a variety sources, there are some clear
trends in the range of observed disk properties.  First, there is an
apparent cutoff at faint surface brightness, which reflects selection
biases against LSBs; the lowest surface brightness galaxies among the
filled circles are barely visible on the photographic plates used for
discovery\footnote{Note that historically only galaxies with
$\mu_0\lta22\surfb$ have been well studied due to their easy
detectability.  While this surface brightness regime spans a
reasonable range in mass, it only covers a limited range in angular
momentum, leading to the (false) impression that, not only do disks
have a characteristic surface brightness, they have a characteristic
angular momentum as well.}.  Second, there is an apparent cutoff at
small exponential scale lengths.  The cutoff results from selection
biases in angular diameter limited surveys from which the galaxies
were drawn.  Angular diameter limited surveys are heavily biased
towards finding galaxies with the largest scale lengths and against
finding intrinsically small galaxies; unless it is incredibly close, a
very small galaxy will not have a large enough angular extent to be
included in the survey.  Therefore, physically large galaxies will be
stongly over-represented in angular diameter limited surveys.

There are two other boundaries in the observed distribution of disk
properties in Figure \ref{alphamufig} which are physically meaningful,
and are not due to selection biases.  First, there is an apparent
absence of large, high surface brightness galaxies, as can be seen
from the diagonal cutoff in the lower right.  This cannot be due to
selection effects, given that these galaxies are the easiest of all
possible galaxies to detect.  However, this cutoff agrees well with
the cutoff in the predicted distribution of scale lengths and surface
brightnesses, calculated above and plotted as the dotted contours.

The second physically important cutoff in the distribution of disk
properties is the absence of high surface brightness galaxies.
Equation \ref{charmueqn} and Figure \ref{alphamufig} suggest that
there will be significant numbers of galaxies which have surface
brightnesses brighter than the Freeman surface brightness (1970).
However, while there are many galaxies with surface brightnesses {\it
fainter} than the Freeman value, there are {\it no} galaxies observed
with surface brightness much brighter than the Freeman value (Allen \&
Shu 1979).

In Figure \ref{alphamufig} there is a maximum surface brightness at
roughly $\mu_0=19.5\surfb$, above which galaxy disks do not seem to
form.  We postulate that these galaxies either lose substantial
amounts of angular momentum during collapse or become unstable to
global modes after formation; both processes would cause the disk to
loose both energy and angular momentum, leading to the formation of
bulges and/or ellipticals.  Support for this idea comes from a
stability analysis of our model disks using the Toomre criteria for
the stability of a rotating stellar disk
($Q\equiv\frac{\sigma_{V_r}\kappa}{3.36G\Sigma}>1$ for stability
(Toomre 1964), where $\sigma_{V_r}$ is the radial velocity dispersion
of the stars, and $\kappa$ is the epicyclic frequency).  While the
Toomre criteria is derived for local disk stability, it gives a
reasonable prediction of a disk's stability to the global modes which
can lead to significant amounts of angular momentum transfer and
dissipation.

In Figure \ref{toomrefig} we plot the Toomre parameter
$Q$ as a function of radius for galaxies with a range of surface
brightnesses.  We have assumed that all of the galaxies have the same
central radial velocity dispersion ($\sigma_{V_r}=100\kms$) and
constant disk thickness, as is implied by observations of edge-on LSBs
(Dalcanton \& Shectman 1996, Cowie et al.\ 1996).  This leads to a
radial dependence on velocity dispersion which falls like
$\exp{r/2\alpha}$, as has been observed for normal galaxies (van der
Kruit \& Freeman 1986, Bottema 1993).  With this choice of
$\sigma_{V_r}$, $Q$'s inverse dependence on surface density leads to
all models with surface densities at or below the Freeman value being
stable, but leads high surface density disks, which are strongly
self-gravitating, to be globally unstable over almost the entire disk.

Given the large uncertainty in the velocity dispersion $\sigma_{V_r}$
outside of the regime of normal galaxies, we have also calculated the
minimum velocity dispersion which would be needed to stabilize the disks
in Figure \ref{toomrefig}.  The high surface density galaxies would
require velocity dispersions which were typically greater than 40\%
of their circular velocity in order to be stable.  However, such
high velocity dispersions would lead to highly non-circular orbits
and thus be likely to produce angular momentum transfer and dissipation.
Thus, we again expect that the high surface density disks would not
take the form predicted by our models.  

These very high density disks would presumably become unstable to bar
formation, which can cause both central concentration and vertical
heating, leading to the formation of a bulge (Combes \& Sanders 1981,
Pfenniger 1984, 1985, Combes et al.\ 1990, Pfenniger \& Norman 1990,
Friedli \& Pfenniger 1990, Pfenniger \& Friedli 1991, Raha et al.\
1991 ).  Although current numerical models for the secular formation
of a bulge find it difficult to make extremely large bulges, these
models usually begin with a ``normal'' disk, as opposed to an
extremely high surface density disk such as the one in Figure
\ref{toomrefig}; we would be most interested in further numerical
studies of the dynamical instabilities of concentrated high surface
density disks.  Observationally, the lack of any significant color
difference between bulges and disks, particularly in the redder bands
($\Delta(R-K)=0.078\pm0.165$; Peletier \& Balcells 1996), as well as
the correlation between disk scale length and bulge scale length (de
Jong 1996, Courteau et al.\ 1996) and preponderance of exponential
bulge profiles (Courteau et al.\ 1996) all support the notion that in
a large fraction of galaxies, the creation of the bulge is closely
tied to the formation of the disk.  Of course, there are likely to be
additional routes to bulge and/or elliptical formation; for a
``messy'' dissipative collapse, the gas may never settle into a disk
at all, and entirely wind up as a condensed elliptical.  Finally,
our stability analysis shows that massive galaxies are more
likely to be Toomre unstable, particularly in the inner regions, which
would suggest that higher mass galaxies should have a larger
bulge-to-disk ratio, in general.  Likewise, galaxies which form from
low angular momenta halos are also more likely to be Toomre unstable,
again leading to a larger bulge-to-disk ratio.  This scenario suggests
that the Hubble sequence might be a sequence of both mass and angular
momentum.

\begin{figure}[hb]
\centerline{ \psfig{figure=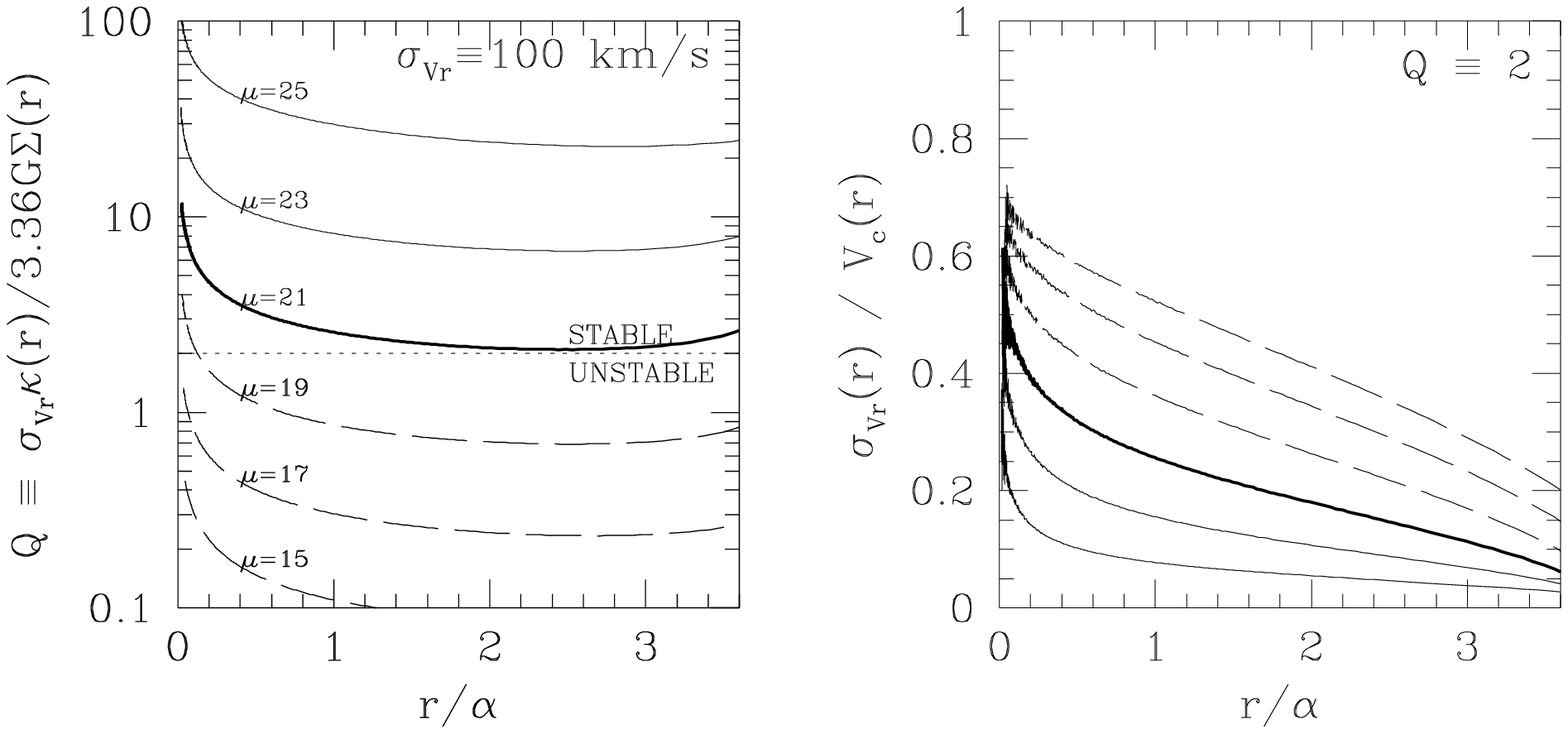,height=4.0in} }
\end{figure}
\figcaption{\label{toomrefig}}
{\footnotesize
\ni
Disk stability as a function of surface brightness.  The left
panel plots the Toomre stability parameter
$Q\equiv\frac{\sigma_{V_r}\kappa}{3.36G\Sigma}$ as a function of
radius (scaled by the exponential scale length $\alpha$), for model
disk galaxies with $\mu_0=15,17,19,21,23,\&25\,B\surfb$ and scale
lengths of $3\kpc$.  The model disk with a characteristic Freeman
value ($\mu_0\sim21$) is plotted as a dark line.  Brighter disks are
plotted as dashed lines, and the LSBs are plotted as solid lines.
The models assume a radial velocity dispersion of $\sigma_{V_r}$
which falls off radially from $100\kmsec$ in the center such that
constant disk thickness is preserved ($\sigma_{V_r}\propto \exp{-r/2\alpha}$).
The high surface brightness disks have $Q<2$ and are likely to be unstable.
The LSBs, which are dark matter dominated, are likely to be 
stable.  
The right panel plots the minimum radial velocity dispersion needed to 
stabilize the disks in the top panel, assuming $Q_{crit}=2$, and plotted
as a fraction of the circular velocity.  The high surface brightness
disks (dashed lines) require radial velocity dispersions which are a
large fraction of their circular velocities in order to be stable to
bar formation and other large scale modes which may result in angular
momentum transfer and secular bulge formation.
}
\clearpage

The predicted distribution of scale lengths and surface brightnesses
in Figure \ref{alphamufig} argues that there should be disks over a
huge range of physical scale and surface brightnesses, well outside of
the range of disk properties represented in large catalogs such as the
NGC and UGC.  There should be large numbers of both intrinsically
small galaxies and low surface brightness galaxies, both of which are
difficult to detect.  While the photographic field surveys which have
produced the largest galaxy catalogs are basically restricted to the
part of the $\alpha$-$\mu_0$ plane occupied by the filled circles,
other specialized surveys can find galaxies in other parts of the
plane.  For example, galaxies in the Local Group (filled triangles)
and from deep surveys of the Virgo cluster are also plotted on Figure
\ref{alphamufig}.  These surveys clearly show that there are indeed
galaxies whose properties lie well outside of the region occupied by
galaxies from larger field surveys.  More importantly, the cluster
surveys plotted in Figure \ref{alphamufig} (open squares and stars) do
not suffer from the strong bias towards intrinsically large galaxies
which affects field surveys, and more closely reflect the intrinsic
distribution of scale lengths; the plotted galaxies from these surveys
are almost entirely found with very small scale lengths, suggesting
that there is indeed the predicted rise towards intrinsically small
galaxies, as shown by the contours of $n(\lg{\alpha},\mu_0)$.  We
expect the rise in the number of physically small galaxies to exist at
both low and high surface brightnesses, suggesting that there may be a
significant but overlooked population of very compact, high surface
brightness galaxies; such galaxies would easily be mistaken for stars,
and may well be as under-represented in large galaxy catalogs as the
population of LSBs.

To make the above comparison between the calculated distribution of
surface density and the observed distribution of surface brightness,
we have assumed that all disks have a single conversion efficiency for
turning gas into stars, such that there is a single disk mass-to-light
ratio $\Upsilon$.  In using a single mass-to-light ratio for all
disks, we have swept a good deal of physics under the rug.  In
particular, we believe that our neglect of detailed physical processes
within the galaxies causes us to systematically overestimate the
surface brightness of LSBs, given that almost all mechanisms for
reducing star formation efficiency are likely to be more effective in
low surface density galaxies than in high surface density ones.
First, if star formation requires a large reservoir of neutral gas,
then the extragalactic ultraviolet background, which will completely
ionize low surface density galaxies (\S\ref{HI}), may suppress or shut
off the formation of stars in these galaxies (Babul \& Rees 1992,
Efstathiou 1992).  Second, if star formation is associated with the
gas in disks becoming locally Toomre unstable to the formation of
spiral structure (implying that the Toomre stability parameter
$Q\propto \sigma\kappa / G\Sigma$ is smaller than some constant, where
$\sigma$ is the velocity dispersion of the gas, $\Sigma$ is the
surface density of the disk, and $\kappa$ is the epicyclic frequency,
set by the disk structural parameters (see Kennicutt 1989)), then one
would expect lower surface density disks to be less unstable for star
formation.  Observations by van der Hulst et al.\ (1993) find that LSB
galaxies do have HI surface densities that that fall below the
critical density implied by $Q$ and are about a factor of 2 lower than
the HI surface densities of HSB galaxies.  Third,
low surface density, low mass galaxies are more likely to lose
their gas through supernova driven winds than high surface density,
high mass galaxies are.  The gas loss both shuts off star formation
prematurely and evolves the galaxy towards larger sizes through sudden
mass loss and subsequent revirialization (Dekel \& Silk 1986, DeYoung
\& Heckman 1994, Babul \& Ferguson 1996) -- both mechanisms which lead
to lower surface brightnesses for low surface density objects.
However, because the gas loss is expected to stop when SN-driven
bubbles ``blow-out'' the top and bottom of the disk, leaving a gas
void whose width is comparable to the thickness of the disk, these
effects are likely more severe in low velocity dispersion halos and
may not play a role in the larger LSB disks.

In spite of these effects, for disks with central surface brightness
brighter than $\mu_0<24\surfb$ (for the range of masses spanned by the
filled circles in Figure \ref{alphamufig}) we have reason to believe
that the assumption of a single value of $\Upsilon$ is valid for the
population as a whole.  As discussed in \S\ref{TFsec}, if $\Upsilon$
varied systematically with surface density over this range, the mean
Tully-Fisher relation in Figure \ref{T-Ffig} would shift up or down
towards brighter or fainter absolute magnitudes, as a function of
surface brightness.  However, because the mean Tully-Fisher relation
is observed to be identical for all disk galaxies with
$\mu_0<24\surfb$ (Zwaan et al.\ 1995, Sprayberry et al.\ 1995), there
cannot be significant variations in $\Upsilon$ with surface
brightness.  Likewise, because the predicted slope agrees closely with
the measured one, there is unlikely to be a systematic variation in
$\Upsilon$ with mass as well.  Zwaan et al.\ (1995) do observe that
the scatter in the Tully-Fisher relationship is about twice as high in
LSBs as in normal galaxies, suggesting that {\it individual} LSB
galaxies are more likely to have values of $\Upsilon$ which depart
from the mean; however, the population as a whole seems to have the
same mean disk mass-to-light ratio.  There may well be systematic
variations in $\Upsilon$ outside of the range of disk properties
spanned by Tully-Fisher measurements; if included theoretically, the
variation in $\Upsilon$ would be manifested in Figure \ref{alphamufig}
as a vertical stretching of the contours of $n(\alpha,\mu_0)$.

\subsection{Selection Effects Upon the Luminosity Function} \label{lumfuncsec}

The predicted distribution of disk scale lengths and surface
brightnesses plotted in Figure \ref{alphamufig} suggests that there
are large numbers of uncataloged galaxies which have not contributed
to measurements of the local luminosity function.  In particular,
although it is rarely discussed explicitly, most galaxy surveys are
sensitive only to galaxies whose central surface brightness exceeds
some minimum critical surface brightness, $\mu_{min}$.  As can be seen
from the contour lines in Figure \ref{alphamufig}, when compared to
the range of surface brightnesses covered by major field surveys
($\mu_0<22\surfb$ typically), the mismeasurement of the luminosity
function may be severe even for bright galaxies.  Furthermore, because
lower mass galaxies also tend to have lower surface brightnesses, the
degree of incompleteness is always worse at faint luminosities; thus a
surface brightness limited sample will always underestimate the faint
end slope of the galaxy luminosity function.

\vfill
\begin{figure}[hb]
\centerline{ \psfig{figure=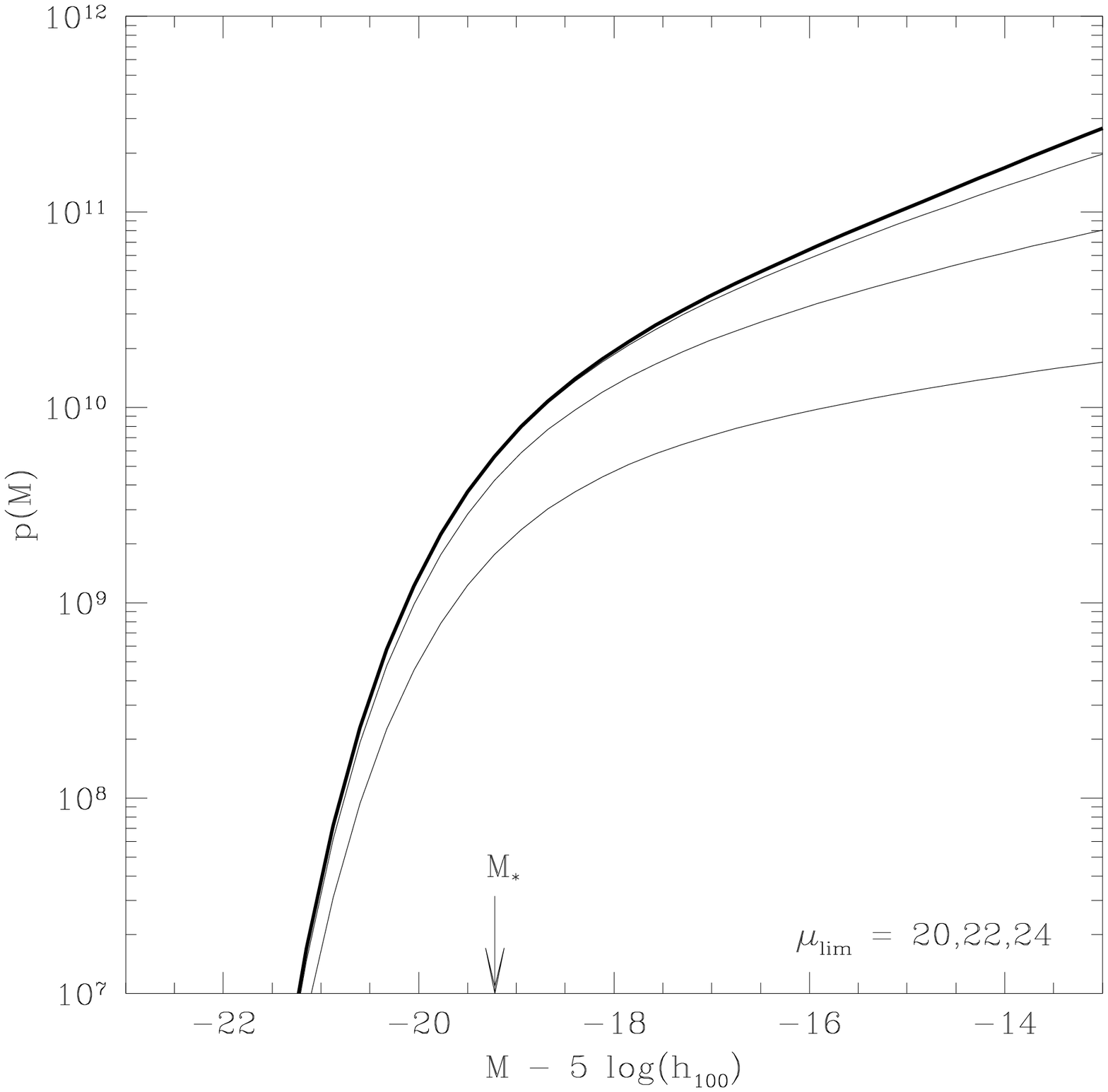,height=4.5in} }
\end{figure}
\figcaption{\label{luminosityfunctionfig}}
{\footnotesize
\ni
The luminosity function which would be observed for different
limiting surface brightnesses (from top to bottom,
$\mu_{lim}=\infty,24,22,20\surfb$), appropriate for a completely unbiased
survey, a deep pencil beam survey, a local field survey, and a
multi-fiber field survey.  We have assumed an intrinsic luminosity
function of the form $p(L)\propto(L/L_*)^{-1.5}\exp(-L/L_*)$, in
arbitrary units of per luminosity per volume.  Note that there is not
only a significant depression of the faint end slope, but also a
potentially large underestimate of the number of bright $L_*$
galaxies.  See Figure \ref{lumdensfig} for the corresponding underestimate
in the luminosity density.  We have assumed the same parameters as in
Figure \ref{alphamufig}.
}
\clearpage

We can quantify the degree to which the limiting surface brightness of
a survey leads to underestimates of the luminosity function.  Equation
\ref{psigMdeqn} gives the distribution of surface brightness at a
fixed disk mass, $M_d$, corresponding to a fixed disk luminosity
$L\equiv M_d/\Upsilon$.  The observed luminosity function is therefore
true luminosity function weighted by the integral of equation
\ref{psigMdeqn} from $0$ to $\mu_{min}$.  Figure
\ref{luminosityfunctionfig} shows the resulting luminosity function
predicted by our model as a function of the surface brightness cutoff,
assuming an underlying luminosity function $n(L)\propto(L/L_*)^{-1.5}
\exp(-L/L_*)$.  The lines correspond to different surface brightness
limits, appropriate to different types of galaxy surveys; shallow
multi-fiber surveys of the field (e.g. LCRS
(Lin 1996)), deeper surveys of the field (e.g. CfA
(Marzke et al 1994)), or deep pencil beam
redshift surveys (e.g. CFRS (Lilly et al.\ 1996)).

First, from Figure \ref{luminosityfunctionfig} it is clear that the
luminosity function can be significantly underestimated, even at
$L_*$.  The field surveys with the shallowest surface brightness
limits can be missing more than a factor of 5 of bright $L_*$
galaxies.  Even deeper surveys of the local field can be missing a
factor of 2.  Only when surveys reach comparable depths to deep pencil
beam surveys are the surveys identifying the majority of bright
galaxies.  This immediately has bearing on the apparent excess of
``faint blue galaxies'' at moderate redshifts; as pointed out by
McGaugh (1994), the surface brightness selection effects
are different in deep pencil beam surveys than in the local field
galaxy surveys, potentially leading to the inclusion of lower surface
brightness galaxies in deep surveys which are otherwised missed in
tabulations of the local galaxy population.  From Figure
\ref{luminosityfunctionfig} it is clear that the expected magnitude of
this effect is rather large, with there being 2-5 times as many bright
galaxies visible in deep surveys as would be expected on the basis of
measurements of the local luminosity function.  We note that this is
remarkably similar to the discrepancy between the observed luminosity
density in deep pencil beam surveys and that which would be derived
from the local field luminosity function (Dalcanton 1993).

Second, there is a systematic decrease in the observed faint end slope
with brighter surface brightness limits; for an intrinsic faint end
slope of $-1.5$, the measured slope would be -1.15, -1.26, or -1.38
for surface brightness limits of $\mu_{min}=20,22,$ or 24.
Semi-analytic models of galaxy formation consistently predict steeper
faint end slopes than are observed, but without making any correction
for surface brightness selection effects (Kauffmann, White \&
Guiderdoni 1993; Lacey et al. 1993; Heyl et al.\ 1994).  While this
shortfall at the faint end has been seen as a failure of theories of
structure formation, such as CDM, Figure \ref{luminosityfunctionfig}
suggests that surface brightness selection effects are easily capable
of producing the shortfall.  We should also note that while
theoretical predictions of the luminosity function do not tend to
overestimate the number of bright galaxies, this could easily be due
to choosing normalizations based upon observations of only the
population of high surface brightness galaxies.

\vfill
\clearpage
\begin{figure}[hb]
\centerline{ \psfig{figure=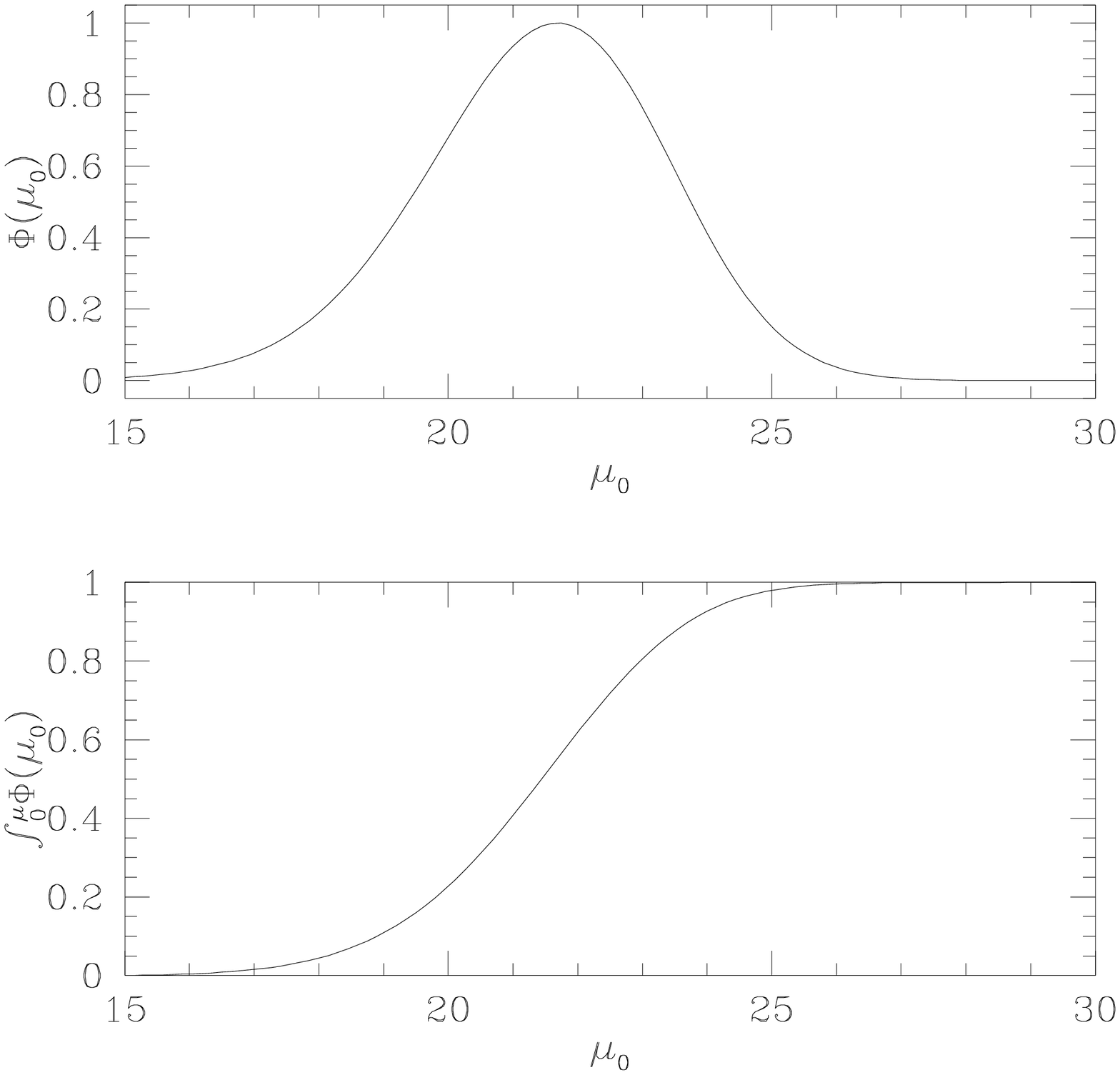,height=6in} }
\end{figure}
\figcaption{\label{lumdensfig}}
{\footnotesize
\ni
The differential and cumulative luminosity density in arbitrary
units as a function of central surface brightness (upper and lower
panels respectively), assuming the same parameters as in Figures
\ref{alphamufig} \& \ref{luminosityfunctionfig}.  Most of the
luminosity density comes from galaxies with central surface
brightnesses between $20$ and $23\surfb$ (upper panel), suggesting
that current surveys are missing a significant fraction of the
luminosity density of the universe.  The lower panel shows that only
deep surveys with $\mu_{lim}=25\surfb$ are capable of detecting most of
the light in the universe.  
}
\clearpage
\vfill
\clearpage
\begin{figure}[hb]
\centerline{ \psfig{figure=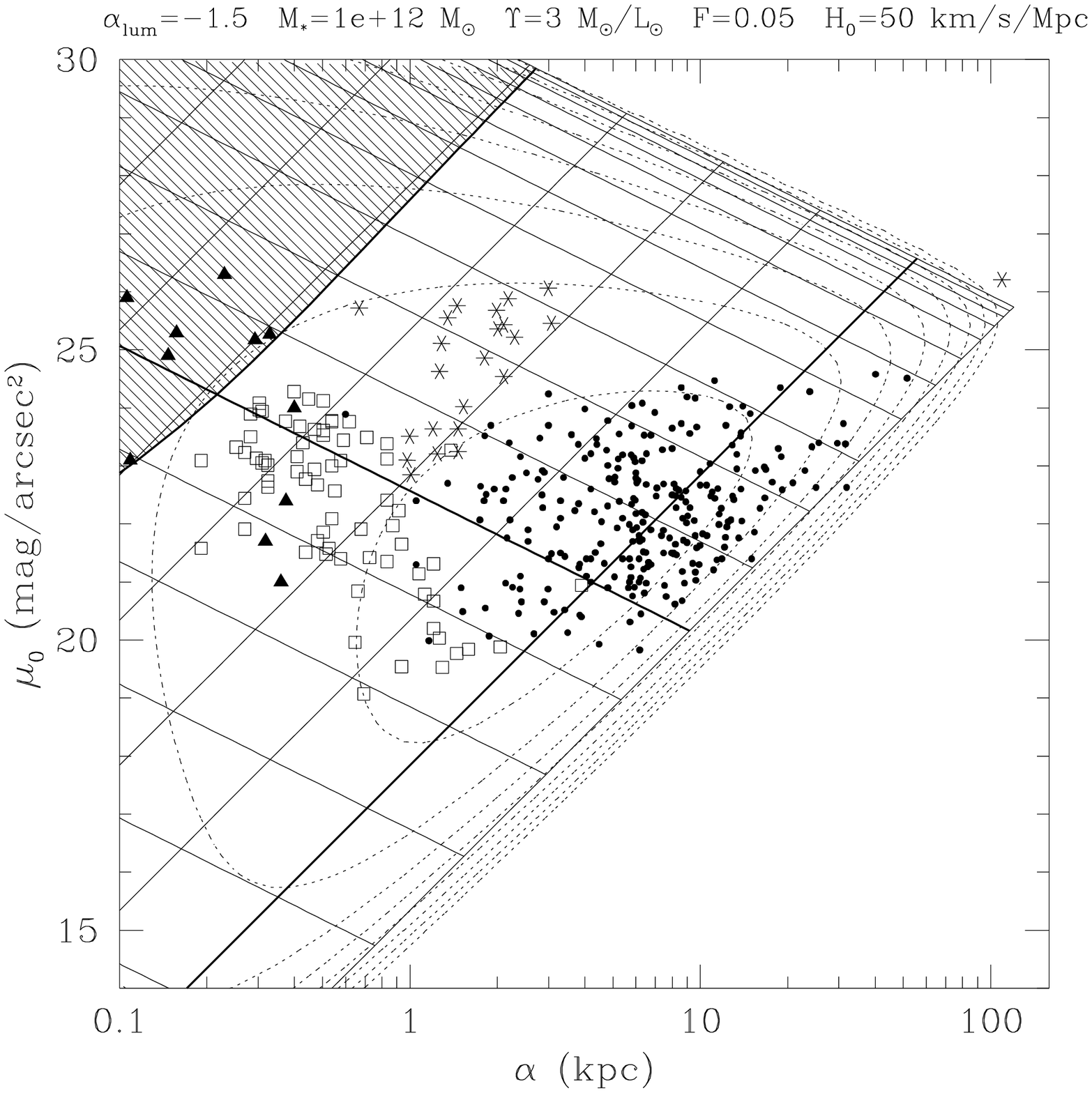,height=6in} }
\end{figure}
\figcaption{\label{almulumdenfig}}
{\footnotesize
\ni
The relative luminosity density as a function of $B$ central
surface brightness $\mu_0$ and scale length $\alpha$.  Identical
to Figure \ref{alphamufig}, except the dotted lines are contours of
constant luminosity density, $\Phi(\lg{\alpha},\mu_0)$, separated
by factors of 10.
}
\clearpage

The potentially large underestimate of the number density of galaxies
suggests that current determinations of the luminosity function may
also underestimate the total luminosity in galaxies.
We can estimate the total luminosity density produced by disc galaxies
as a function of surface brightness by integrating the intrinsic luminosity
function weighted by $p(\mu_0|L)$ (equation \ref{psigMdeqn}) over 
luminosity:

\begin{equation}				 \label{lumdenseq}
\phi(\mu_0) = \int_{L_{min}}^\infty n(L)p(\mu_0|L)LdL,
\end{equation}

\ni where $L_{min}\equiv M_{min}/\Upsilon$, from equation \ref{Mmineqn},
is the expected cutoff in the galaxy luminosity function.
Figure \ref{lumdensfig} shows the resulting luminosity density as a function
of surface brightness, assuming a faint end slope of
$\alpha_{lum}=-1.5$.  The lower panel in the figure shows the
cumulative luminosity density distribution; it suggests that
only deep surveys with $\mu_{lim}\approx24\surfb$ are capable of
observing most of the luminosity density of the universe
(see, e.g., Vaisanen 1996).  Figure \ref{almulumdenfig} shows
how the relative contribution to the luminosity density changes
as a function of galaxy surface brightness and scale length.
From the contours of constant luminosity density, it is clear
that galaxies with central surface brightness as faint as
$25\surfb$ and with scale lengths as small as $0.2\kpc$ make
significant contributions to the luminosity density of the
universe.

The underestimate of the local luminosity function has strong
implications for measurements of the mass density of the universe,
$\rho_0$.  One common technique to measure $\rho_0$ is to take the
local luminosity density and scale it by the large scale mass-to-light
ratio measured in rich clusters such as Coma.  However, if the
luminosity of Coma were to be determined over a much wider surface
brightness range than the local luminosity function, then the
appropriate mass-to-light ratio for calculating $\rho_0$ would be much
larger than the measured mass-to-light ratio for Coma.  Failing to
correct for this effect could lead to measured values of $\rho_0$
which are factors of 2 or more too low.

\section{Gas Detection}				\label{HI}

We have postulated that hierarchical structure formation models
naturally lead to a large population of low surface brightness disk
galaxies.  While such a pervasive population of LSB's is difficult to
detect optically, it could potentially manifest itself in blind HI
surveys.  However, it has often been considered a failing of
hierarchical structure formation scenarios that deep HI surveys have
failed to uncover a significant population of dwarf galaxies.  The few
uncataloged dwarfs that are discovered are preferentially found near
bright galaxies (see van Gorkom 1993 for a recent review).  There does
not seem to be a large population of gas rich dwarfs in the small
volume which has been explored in blind HI surveys.

However, in light of recent work showing a sharp cutoff in HI disks at
column densities of $10^{19}\cm^{-1}$ ( Corbelli, Schneider, \&
Salpeter 1989), the paucity of HI dwarfs may not be surprising.
Recent work by Maloney (1993) and Corbelli \& Salpeter (1993)
convincingly demonstrates that ionization of the HI by a UV background
accounts for the sharp cutoff in HI disks.  As we have shown that
low-mass galaxies tend to have low surface densities, these galaxies
will be prone to having their hydrogen ionized, reducing their
detectable HI masses well below their total hydrogen masses.  On the
other hand, HI observations of LSBs find that while the HI surface
density of LSBs are less than normal galaxies, the ratio of HI mass to
blue luminosity is high, suggesting that LSBs are in general less
evolved than normal galaxies, due either to formation time, or to
reduced star formation efficiency (van der Hulst et al.\ 1993, de Blok
et al.\ 1996).  This effect may partially counteract the likely
decrease in the neutral fraction due to ionization.  However, the
ratio of HI mass to blue luminosity is only a factor of $\approx2$
higher than normal galaxies, while the surface brightnesses are
factors of ten lower, suggesting that the two competing effects
are unlikely to cancel completely.

Low surface density galaxies are therefore likely to suffer from
strong biases against their detection in HI surveys, not just in
optical surveys.  A large population of LSB's could easily have been
overlooked by existing surveys.  The dwarfs that have been detected in
HI surveys must have higher surface densities in general, and thus are
more likely to have collapsed earlier from larger overdensities.  This
would explain why these dwarfs are found to trace the bright galaxy
population.  (For a more complete discussion of constraints on the LSB
population from $21\cm$ surveys, see McGaugh (1996).)  However, while
the neutral gas fraction of LSBs may be small enough to keep them out
of surveys of HI emission, they may well be seen in absorption,
particularly in the Lyman-$\alpha$ forest.

\section{Summary}		\label{generalprops}

In this paper, we have developed a scenario that links the mass and angular
momentum of a proto-galaxy to the luminosity and surface brightness
of the resultant galactic disk.  Gas in low angular momentum proto-galaxies
collapse by large factors to form high surface brightness galaxies.
Low surface brightness galaxies form from low mass and high
angular momentum proto-galaxies.

As gravitational collapse in any hierarchical model
with Gaussian initial conditions leads to a broad
distribution of halo masses and angular momentum,
we expect that galaxies should have a wide range of surface
brightness and disk scale lengths.  We expect that some disk
galaxies form with surface brightnesses less than
 $27\,B\surfb$, well
below current observational limits.  We also expect the number density
of galaxies to rise towards small scale lengths and low surface
brightnesses, as is seen in surveys of nearby clusters.  This
correlation also leads to an apparent correlation between surface
brightness and magnitude, as is observed (Bingelli et al.\ 1984,
Ferguson \& Sandage 1988).

Following Mestel (1963), Fall \& Efstathiou (1980), Gunn
(1982), van der Kruit (1987), we follow the collapse of
a uniform gas cloud that has experienced a uniform external torque.
We extend the earlier work by calculating the collapse
in a realistic dark matter halo and use the adiabatic
approximation developed by Blumenthal et al. (1986) to model the
response of the halo to the disk.
This scenario produces galaxy disks with 
asymptotically flat rotation curves and exponential light profiles,
where the disk scale length is roughly proportional
to $\lambda M^{1/3}$.  Here, $\lambda$ is the spin parameter
of the proto-galaxy
and $M$ is its mass. This relation leads to galaxies
that obey the Tully-Fisher law  as
$v_c^2 \propto M(r_d)/r_d \propto M^{2/3} \propto L^{2/3}$.
It also helps explain
the disk-halo conspiracy (Bahcall \& Casertano 1985) associated
with featureless rotation curves  as both the disk and
halo scale lengths are proportional to $M^{1/3}$.

The shape of the rotation curve depends upon a galaxy's 
angular momentum.  Low angular momentum
disks are centrally concentrated and dominate the inner
portions of the rotation curve.  Thus, these galaxies have rapidly
rising rotation curves in the centers.
High angular momentum disks have their mass spread
out to larger radii, leading to a smaller dynamical contribution from
the disk, relative to the halo, at all radii.  Thus, high angular
momentum disks, which tend to be low surface brightness, have more
slowly rising rotation curves.
Low surface brightness disks also have apparently higher
dynamical mass-to-light ratios, because their extended disks
encompass a larger
fraction of the dark matter halo within their optical radius.
Because the disk makes a relatively small contribution
to the galactic potential in LSBs, it serves as an
excellent tracer of the initial dark mass profile and the shape
of the dark halo.
Furthermore, at a given mass, low surface brightness galaxies have
much larger scale lengths, allowing them to probe the halo at much
larger radii.

This model predicts that very low angular momentum proto-galaxies
collapse to form very high surface brightness galaxies.  Since
these very high surface brightness galaxies are globally
unstable to non-axisymmetric perturbations, they likely form
bars, bulges and/or ellipticals.  Thus, the Toomre instability
criterion may explain why few disk galaxies have  central
surface brightnesses above the canonical Freeman value.

Our model
suggests that current galaxy surveys are missing a 
significant fraction of the total number
of galaxies, and have severely underestimated the faint end slope, due to
surface brightness limits which prevent them from finding low
surface brightness galaxies.  The luminosity density of the universe
is similarly underestimated.  The underestimate should be much less in
deep pencil beam redshift surveys, and is of the right magnitude to
explain much of the apparent ``excess'' in the galaxy population at
moderate redshifts.  It is intriguing that wide-field faint
searches for LSBs (Dalcanton et al. 1997) are finding the
large number of LSBs predicted by this model.

The galaxy formation scenario makes  a
number of assumptions  that require further
observational and numerical testing, namely: (1) that gravitational
collapse leads to a universal dark matter halo (Navarro et al.\ 1996);
(2) that angular momentum is conserved through galaxy collapse and
(3) that the initial angular momentum distribution is similar to that
produced by a external tidal torque.  All three of these
assumptions are controversial.  However, wherever possible, we have
shown that our results are well supported by existing observations.
The success of our model in matching the observations
suggests that, in spite of the simplicity of our assumptions, the
overall scenario merits careful examination.

\bigskip
\bigskip
\centerline{Acknowledgements}
\medskip

We thank Arif Babul, John Bahcall,
Daniel Eisenstein, Jim Gunn, and Ian Smail for
usefule discussions, and the referee, P.\ C.\ van der Kruit for
very helpful suggestions.  DNS was supported by NASA. Support for JJD was
provided by NASA through Hubble Fellowship grant \#2-6649 awarded by
the Space Telescope Science Institute, which is operated by the
Association of Universities for Research in Astronomy, Inc., for NASA
under contract NAS 5-26555.

 \vfill

\section{References}

\hi{Allen, R. J., \& Shu, F. H. 1979, \apj 227, 67.}

\hi{Athanassoula, E. 1984, Physics Reports, 114, 321.}

\hi{Babul, A. 1989, Ph.D. thesis, Princeton University.}

\hi{Babul, A., \& Ferguson, H. C. 1996, \apj 458, 100.}

\hi{Babul, A., \& Rees M. J. 1992, \mn 255, 346.}

\hi{Bahcall, J.N. \& Casertano, S. 1985, \apj 293, L7.}

\hi{Barnes, J. E., \& Efstathiou, G. 1987, \apj 319, 575.}

\hi{Bender, R., Burstein, D. \& Faber, S.M. 1992, \apj 399, 462.}

\hi{Bernstein, G. M., Tyson, J. A., Brown, W. R., \& Jarvis, J. F. 1993,
\apj 426, 516.}

\hi{Bernstein, G.M., et al. 1994, \aj 107, 1962.}

\hi{Bingelli, B., Sandage, A., \& Tarenghi, M. 1984, \aj 89, 64.}

\hi{Blumenthal, G. R., Faber, SM, Flores, R \& Primack, JR  1986 \apj 301, 27.}

\hi{Bond, J. R., Cole, S., Efstathiour, G. \& Kaiser, N. 1991, \apj 379, 440.}

\hi{Boroson, T. 1981, \apjs 46, 177.}

\hi{Bothun, G. D., Beers, T. C., Mould, J. R., \& Huchra, J. P. 1986,
\apj 308, 510.}

\hi{Bothun, G. D., Impey, C. D., \& Malin, D. F. 1991, \apj 376, 404.}

\hi{Bothun, G. D., Impey, C. D., Malin, D. F., \& Mould, J. 1987, \aj 94, 23.}

\hi{Bothun, G. D., Schombert, J. M., Impey, C. D., Sprayberry, D., \&
McGaugh, S. S. 1993, \aj 106, 530.}

\hi{Bottema, R. 1993, \aa 275, 16.}

\hi{Bower, R. J. 1991, \mn 248, 332.}

\hi{Burstein, D., \& Rubin, V. C. 1985, \apj 297, 423.}

\hi{Caldwell, N., Armandroff, T. E., Seitzer, P., \& Da Costa, G. S., 1992
\aj 103, 840.}

\hi{Catelan, P. \& Theuns, T. 1996a, to appear in \mn.}

\hi{Catelan, P. \& Theuns, T. 1996b, astro-ph/9604078.}

\hi{Cayatte, V., Kotanyl, C., Balkowski, C., \& van Gorkom, J. H. 1994,
\aj 107, 1003.}

\hi{Combes, F. \& Sanders, R. H. 1981, \aa 96, 164.}

\hi{Combes, F., Debbasch, F., Friedli, D., \& Pfenniger, D. 1990, \aa 233, 82.}

\hi{Corbelli, E. \& Salpeter, E 1993, \apj 419, 104.}

\hi{Corbelli, E., Schneider, S. E., \& Salpeter, E. 1989, \aj 97, 390.}

\hi{Courteau, S., de Jong, R. S., \& Broeils, A. H. \apjl 457, L73.}

\hi{Cowie, L, L., Hu, E. M., \& Songaila, A., 1996, \aj submitted (CHS).}

\hi{Dalcanton, J. J. 1993, \apjl 415, L87.}

\hi{Dalcanton, J. J. 1995, Ph.D. Thesis, Princeton University.}

\hi{Dalcanton, J. J., Spergel, D. N., 
Gunn, J. E., Schmidt, M., \& Schneider, D. P. 1997 \aj, submitted.}

\hi{Dalcanton, J. J. \& Shectman, S. A. 1996, \apjl 465, L9.}

\hi{Davis, M., Efstathiou, G., Frenk, C., \& White, S. 1985, \apj 292, 371.}

\hi{Davies, J. I., Phillipps, S., \& Disney, M. J. 1988, \mn 231, 69P.}

\hi{Davies, J. I., Phillipps, S., \& Disney, M. J. 1990, \mn 244, 385.}

\hi{de Blok, W. J. G., van der Hulst, J. M., \& Bothun, G. D. 1995, \mn 274,
235.}

\hi{de Blok, W. J. G., McGaugh, S. S., \& van der Hulst, J. M. 1996, \mn,
in press.}

\hi{de Jong, R. S. 1995, Ph.D.\ Thesis, University of Groningen.}

\hi{de Jong, R. S., \& van der Kruit 1994, \aasup 106, 451.}

\hi{de Jong, R. S., 1996, \aasup, 118, 557.}

\hi{Dekel, A., \& Silk, J. 1986, \apj 303, 39.}

\hi{DeYoung, D., \& Heckman, T. 1994, \apjl 431, 598.}

\hi{Disney, M. 1976, \nat 263, 573.}

\hi{Dubinski, J.  \& Carlberg, R.G. 1991, ApJ 378, 496.}

\hi{Eisenstein, D. J., \& Loeb, A. 1995, \apj  439, 520.}

\hi{Eisenstein, D. J. 1996, Harvard Ph.D. Thesis.}

\hi{Efstathiou, G. 1992, \mn 256, 43p.}

\hi{Efstathiou, G. \& Silk, J. 1983, \fcp 9, 1.}

\hi{Efstathiou, G., Bernstein, G., Katz, N., Tyson, J. A., \& Guhathakurta, P.
1991, \apjl 380, L47.}

\hi{Evrard, A. E., Summers, F. J., \& Davis, M. 1994, \apj 422, 11.}
 
\hi{Faber, S. M. 1982, in {\rm Astrophysical Cosmology}, ed. H. A. Bruck,
G. V. Coyne, \& M. S. Longair, (Vatican: Pontificia Acadamia Scientiarum),
191.}

\hi{Ferguson, H. C., \& McGaugh S. S. 1995, \apj 440, 470.}

\hi{Ferguson, H. C., \& Sandage, A. 1988, \aj 96. 1520.}

\hi{Flores, R., Primack J. R., Blumenthal, G. R., \& Faber, S. M.
1993, \apj 412, 443.}

\hi{Freeman, K. C. 1970, \apj 160, 811.}

\hi{Gallagher, J. S.,  \& Hunter, D. A. 1984, \araa 22, 37.}

\hi{Gnedin, O., Goodman, J., \& Frei, Z. 1985, \aj 110, 1105.}

\hi{Goad, J. W., \& Roberts, M. S. 1981, \apj, 250, 79.}

\hi{Gunn, J. E. 1982, in {\rm Astrophysical Cosmology}, ed. H. A. Bruck,
G. V. Coyne, \& M. S. Longair, (Vatican: Pontificia Acadamia Scientiarum),
191.}

\hi{Heyl, J. S., Cole, S., Frenk, C. S., \& Navarro, J. F. 1994, astro-ph/9408065.}

\hi{Hoffman, Y., Silk, J., Wyse, R. F. G. 1992, \apjl 388, L13.}

\hi{Impey, C. D., Bothun, G. D., \& Malin, D. F. 1988, \apj 330, 634.}

\hi{Impey, C. D., Sprayberry, D., Irwin, M., \& Bothun, G. D. 1996, \apjs 105,
209.}

\hi{Irwin, M. J., Davies J. I., Disney, M. J., \& Phillips, S. 1990, \mn
245, 289.}

\hi{Jeans, J. H. 1929, {\rm Astronomy and Cosmogony,} 2nd ed., (Cambridge,
England: Cambridge University Press).}

\hi{Kaiser, N. 1984, \apjl 284, L9.}

\hi{Katz, N., \& Gunn, J. E. 1991 \apj 377, 365.}

\hi{Kauffmann, G., White, S.D.M. \& Guiderdoni, B. 1993, \mn 264, 201.}

\hi{Kennicutt, R. C. 1989, \apj 344, 685.}

\hi{Koo, D. C., \& Szalay, A. 1984, \aj 282, 390.}

\hi{Knezek, P. 1993, Ph.D. Thesis, University of Massachusetts.}

\hi{Lacey, C., \& Cole, S. 1993, \mn 262, 627.}

\hi{Lacey, C., Guiderdoni, B., Rocca-Volmerange, B. \& Silk, J.
1993, \apj, 402, 15.}

\hi{Lilly, S. J., Tresse, L., Hammer, F., Crampton, D., \& Le Fevre, O.
1995, \apj 455, 108.}

\hi{Lin, H., Kirshner, R. P., Shectman, S. A., Landy S. D., Oemler, A.,
Tucker, D. L., \& Schechter, P. L. 1996, \apj submitted.}

\hi{Lo, K. Y., Sargent, W. L. W., \& Young, K. 1993, \aj 106, 507.}

\hi{Loveday, J., Peterson, B. A., Efstathious, G., Maddox, S. J. 1992, \apj
390, 338.}

\hi{Maloney, P. 1993, \apj, 414, 41.}

\hi{McGaugh, S. S., \& Bothun, G. D. 1994, \aj 107, 530.}

\hi{McGaugh, S. S. 1994, \apj 426, 135.}

\hi{McGaugh, S. S. 1994, \nat 367, 538.}

\hi{McGaugh, S. S. 1996, \mn 280, 337.}

\hi{McGaugh, S. S., Schombert, J. M., \& Bothun, G. D. 1995, \aj 109, 2019.}

\hi{Mestel, L.  1963 \mn 126, 553.}

\hi{Mo, H. J., McGaugh, S. S., \& Bothun, G. D. 1994, \mn 267, 129.}

\hi{Mo, H. J. \& White, S. D. M. 1995, submitted to \mn, astro-ph/9512127.}

\hi{Navarro, J.F., Frenk, C.S. \& White, S.D.M. 1996, \apj 462, 563.}

\hi{Navarro, J.F., \& Steinmetz, M., 1996, astro-ph/9605043.}

\hi{Navarro, J.F., \& White, S.D.M., 1993, \mn 265, 271}
 
\hi{Nilson, P. 1973, Uppsala General Catalog of Galaxies, {\rm Uppsala
Astr. Obs. Ann.}, Vol.6.}

\hi{Ostriker, J. P. \& Peebles, P. J. E.  1973, \apj, 186, 487.}	

\hi{Peacock, J. \& Heavens A. 1989, \mn 243, 133.}

\hi{Peebles, P. J. E. 1969, \apj 155, 393.}

\hi{Persic, M., \& Salucci, P. 1995, \apjs 99, 501.}

\hi{Persic, M., Salucci, P., \& Stel, F. 1996, \mn 281. 27.}

\hi{Pfenniger D. 1984, \aa 134, 373.}

\hi{Pfenniger D. 1985, \aa 150, 112.}

\hi{Pfenniger D., \& Norman, C. 1990, \apj 363, 391.}

\hi{Pfenniger D., \& Friedli, D. 1991, \aa 252, 75.}

\hi{Phillipps, S., Davies, J. I., \& Disney, M. J. 1990, \mn 242, 235.}

\hi{Press, W. H., \& Schechter, P. L. 1974, \apj 330, 579.}

\hi{Pritchett C. J., \& Infante L. 1992, \apjl 399, L35.}

\hi{Quinn, T. \& Binney, J. 1992, MNRAS, 255, 729.}

\hi{Raha N., Sellwood, J. A., James, R. A., \& Kahn, F. D. 1991, \nat
352, 411.}

\hi{Ryden, B.S. 1988, ApJ, 329, 589.}

\hi{Romanishin, W., Strom, K. M., \& Strom, S. E. 1983, \apjs 53, 105.}

\hi{Schechter, P. 1975, Caltech Ph.D. Thesis.}

\hi{Schombert, J. M., \& Bothun, G. D. 1988, \aj 95, 1389.}

\hi{Schombert, J. M., Bothun, G. D., Schneider, S. E., \& McGaugh, S. S. 1992,
\aj 103, 1107}

\hi{Shapiro, P. R., Giroux, M. L., \& Babul, A. 1994, \apj 427, 25.}

\hi{Shectman, S. A. 1974, \apj 188, 233.}

\hi{Sprayberry, D., Bernstein, G. M., Impey, C. D., \& Bothun, G. D. 1995a, \apj 438, 72.}

\hi{Sprayberry, D., Impey, C. D., Bothun, G. D., Irwin, M. J. 1995b, \aj 109, 558.}

\hi{Sprayberry, D., Impey, C. D., \& Irwin, M.J. 1996, \apj 463, 535}

\hi{Stevenson, P. R. F., Shanks, T., Fong, R., \& MacGillivray, H. T. 1985,
\mn 213, 953.}

\hi{Strauss, M. A., \& Willick, J. A. 1995, Phys.\ Rep.\ 261, 271.}

\hi{Toomre, A. 1964, \apj 139, 1217.}

\hi{Tully, R. B., \& Fisher J. R. 1977, \aa 54, 661.}

\hi{Turner, J. A., Phillips, S., Davies, J. I., \& Disney, M. J. 1993, \mn
261, 39.}

\hi{Vaisanen, P. 1996, CfA preprint 4308 (astro-ph/9604147), to appear in \aa}

\hi{van der Hulst, J. M., Skillman, E. D., Smith, T. R., Bothun, G. D.,
McGaugh, S. S., \& de Blok, W. J. G. 1993, \aj 106, 548.}

\hi{van der Kruit, P. C. 1987 AA 173, 59.}

\hi{van der Kruit, P. C. \& Freeman, K. C. 1986, \apj 303, 556.}

\hi{Warren, M. S., Quinn, P. J., Salmon, J. K., \& Zurek, W. H. 1992, \apj
399, 405.}

\hi{White, S. D. M. 1994, MAP preprint.}

\hi{White, S. D. M., David, M., Efstathiou, G., \& Frenk, C. S. 1987, \nat
330, 451.}

\hi{White, S. D. M., \& Frenk, C. S. 1991, \apj 379, 52.}

\hi{Willick, J. A., Courteau, S., Faber, S. M., Burstein,  D., Dekel, A.,
\& Kolatt, T. 1996, \apj 457, 460.}

\hi{Willick, J. A., Courteau, S., Faber, S. M., Burstein,  D., \& Dekel, A.
1995, \apj 446, 12.}

\hi{Zhang, X. 1996, \apj 457, 125.}

\hi{Zwaan, M. A., van der Hulst, J. M., de Blok, W. J. G., \& McGaugh, S. S
1995, \mn 273, L35.}

\hi{Zwicky, F. 1957, in {\rm Morphological Astronomy}, (New York:
Springer-Verlag).}

\vfill
\clearpage

\section{Appendix}

In this appendix, we redo the calculation presented in the main text, but
here model the initial halo density distribution as $\rho (r)=B\rho
_0r_0/[r_i(1+r_i^2/r_0^2)]$. This density distribution is close to that fit
by Navarro, Frenk and White (1996) to their numerical simulations. In their
simulations, $B\simeq 7\times 10^4$, and the density profile extends to $
r_{\max }\simeq 7.4r_0$. The mass profile that corresponds to the initial
dark halo mass distribution can be inverted:

\begin{equation}
r_i=r_0\sqrt{\left( 1+\frac{r_{\max }^2}{r_0^2}\right) ^{m_h}-1}
\end{equation}

As in section 2, we can calculate the final radial distribution of the disk
and the halo by requiring angular momentum conservation:

\begin{equation}
m_d=1-\left[ 1-\beta m_h^{1/2}\sqrt{\left( 1+\frac{r_{\max }^2}{r_0^2}%
\right) ^{m_h}-1}\right] ^{3/2}
\end{equation}

\begin{equation}
r_f=\frac{m_hr_0\sqrt{\left( 1+\frac{r_{\max }^2}{r_0^2}\right) ^{m_h}-1}}{%
(1-F)m_h+Fm_d}
\end{equation}

\ni where $\beta =\sqrt{G(1-F)M_{tot}}/J_{\max }$.  For $r_{max} = 7.4
r_0$, the energy of the halo is $-0.085 GM_{tot}^2/r_0$, thus, $r_0 =
0.141 a$, where $a$ is the radius of the initial fluctuation that
collapsed to form the galaxy.  Hence, $\beta \simeq
0.15/\lambda$. Figure \ref{appendixsigmafig} shows the rotation curves
and surface density profiles for a range of values for $\lambda $ and
$F$. For this model, the disk scale length, $r_{disk}\simeq a\lambda
^br_0$, where $a = 4.6*(1+20 f)$ and $b = 1+6f$

Following the same approach used in sections 2 and 3, we can compute the
predicted distribution of surface densities at a given luminosity:

\begin{equation}
p(\mu _0|L)=\frac \delta {1.08b\sqrt{2\pi }}\exp \left[ -\frac{(\mu _0-%
\overline{\mu }(L))^2}{2(1.08\delta )^2b^2}\right] d\mu _0
\end{equation}

\ni where 

\begin{equation}
\overline{\mu }(L)=-2.5\log _{10}\left[ \frac{B^2}{2\pi a^2}\left( \frac{%
F\rho _0}\Upsilon \right) ^{2/3}\frac{L^{1/3}}{\langle \lambda \rangle ^{2b}}%
\right] =\mu _{*}+\frac 56\log _{10}\left( \frac L{L_{*}}\right) 
\end{equation}

\ni This model yields similar predictions to the Hernquist model used in the
main text; however, now the width of the central surface density
distribution is slightly broader: $\sigma _{disk}=1.08\delta (1+6f)$.

\begin{figure}[hb]
\centerline{ \psfig{figure=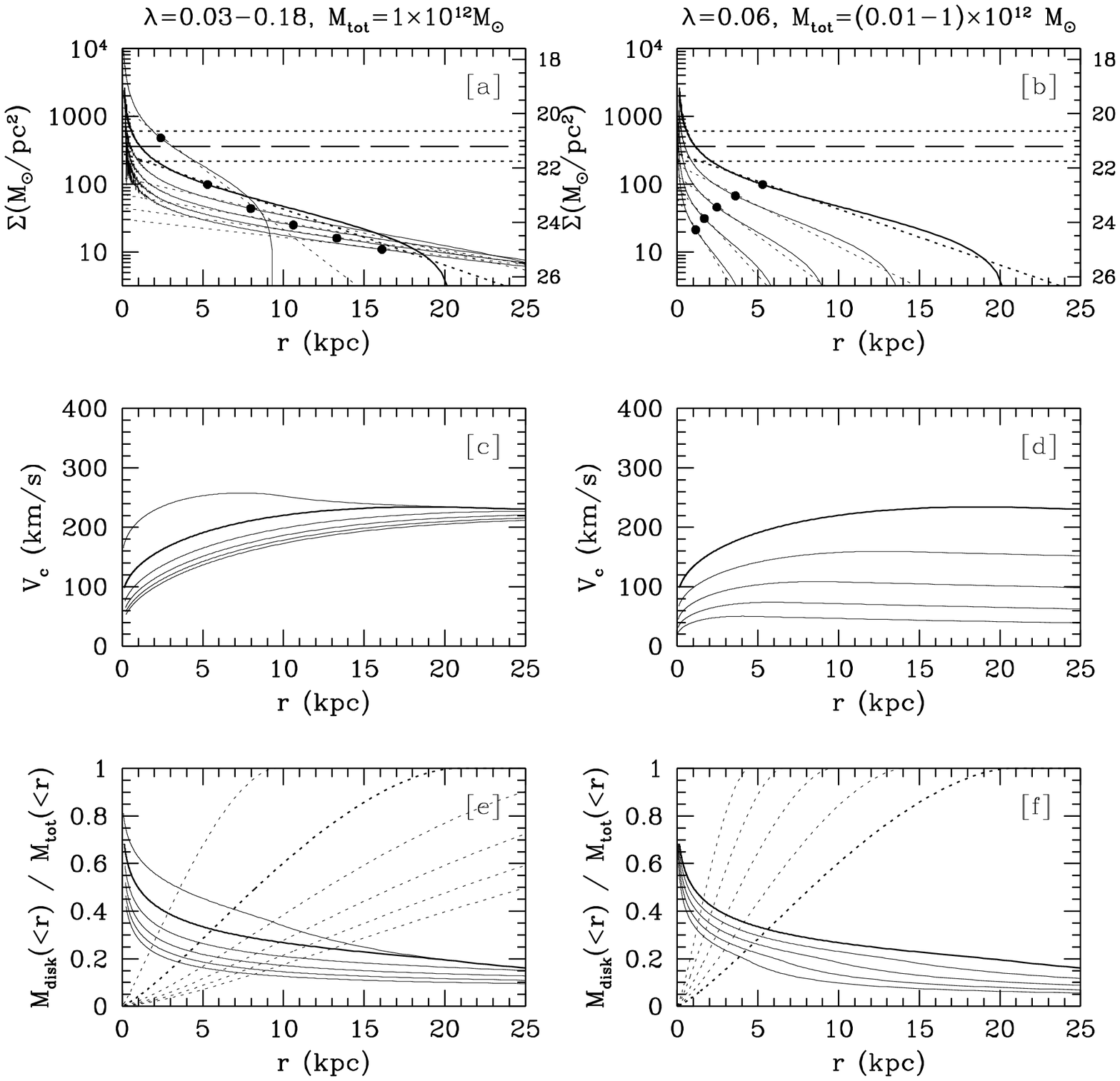,height=6in} }
\end{figure}
\figcaption{\label{appendixsigmafig}}
{
\footnotesize
\ni
Same as Figure \ref{surfbrightfig}, but for an initial halo density
profile of the form: $\rho(r_i) \propto r_i^{-1} (r_i^2 + r_0^2)^{-1}$
}
\vfill
\clearpage

\end{document}